\documentclass[11pt,a4paper]{article}
\usepackage{amsmath, amsfonts,amssymb,mathtools,nicefrac,nccmath,cases}
\usepackage{microtype,booktabs,multirow}
\usepackage[usenames,dvipsnames,table]{xcolor}
\usepackage{colortbl}
\usepackage{tikz}
\usetikzlibrary{shapes,arrows}
\usetikzlibrary{calc,shapes.callouts,shapes.arrows,bending,intersections}
\usetikzlibrary{shapes.geometric}
\usetikzlibrary{arrows.meta,arrows}
\usetikzlibrary{decorations.pathmorphing}
\usetikzlibrary{decorations.markings}
\usetikzlibrary{shapes.multipart}
\usetikzlibrary{tikzmark}
\usepackage{arydshln}
\usepackage{caption}
\usepackage{subcaption}
\usepackage{amssymb}
\usepackage{amsmath}
\usepackage{physics}
\usepackage{amsmath}
\usepackage{tikz}
\usepackage{mathdots}
\usepackage{yhmath}
\usepackage{cancel}
\usepackage{color}
\usepackage{siunitx}
\usepackage{array}
\usepackage{multirow}
\usepackage{amssymb}
\usepackage{gensymb}
\usepackage{tabularx}
\usepackage{booktabs}
\usetikzlibrary{patterns}
\usetikzlibrary{shadows.blur}
\usetikzlibrary{shapes}
\usepackage{float}
\usetikzlibrary{fadings}
\usetikzlibrary{patterns}
\usetikzlibrary{shadows.blur}
\usetikzlibrary{shapes}
\usepackage[nosort]{cite}
\usepackage{colortbl}

\usepackage{xcolor}
\usepackage{lipsum}

\usepackage[linktocpage=true]{hyperref}
\hypersetup{colorlinks=true,linkcolor=nicecolor,citecolor=nicecolor,urlcolor=nicecolor}
\definecolor{nicecolor}{rgb}{0.1, 0.3, 0.4}

\definecolor{blue}{rgb}{0.06, 0.3, 0.57}
\definecolor{Gray}{gray}{0.4}

\definecolor{nicecolor}{rgb}{0.1, 0.3, 0.4}
\definecolor{blue}{rgb}{0.06, 0.3, 0.57}
\definecolor{Gray}{gray}{0.4}
\colorlet{tableheadcolor}{gray!15} 
\colorlet{tablerowcolor}{gray!7} 

\def\hybrid{\topmargin -20pt    \oddsidemargin 0pt
	\headheight 0pt \headsep 0pt
	\textwidth 6.5in        
	\textheight 9in         
	\textwidth 6.25in       
	\textheight 9 in       
	\marginparwidth .875in
	\parskip 5pt plus 1pt 
	\jot = 1.5ex
}
\hybrid
\numberwithin{equation}{section}
\numberwithin{table}{section}
\usepackage{float}
\usepackage{tabularx}
\newcolumntype{D}{>{\centering\arraybackslash}X}
\newcolumntype{L}{>{$}l<{$}}
\newcolumntype{R}{>{$}r<{$}}
\newcolumntype{C}{>{$}c<{$}}
\usepackage{IEEEtrantools}
\usepackage{makecell}

\newcommand{\beq}{\begin{equation}}  \newcommand{\eeq}{\end{equation}}
\newcommand{\bal}{\begin{aligned}}   \newcommand{\eal}{\end{aligned}}
\newcommand{\bea}{\begin{eqnarray}}  \newcommand{\eea}{\end{eqnarray}}
\def\beqa{\begin{eqnarray}}
\def\eeqa{\end{eqnarray}}

\newcommand{\bmat}{\left(\begin{array}}
\newcommand{\emat}{\end{array}\right)}

\usepackage{cancel}

\newcommand{\be}{\begin{equation}}
\newcommand{\ee}{\end{equation}}

\definecolor{Gray}{gray}{0.95}

\begin{document}

\baselineskip=14pt
\parskip 5pt plus 1pt

\vspace*{-1.5cm}
\begin{flushright}    
  {\small 
  
  }
\end{flushright}

\vspace{2cm}
\begin{center}        

  {\huge de Sitter Bubbles and the Swampland\\
   [.3cm]  }
 
\end{center}

\vspace{0.5cm}
\begin{center}        
{\large  Alek Bedroya, Miguel Montero, Cumrun Vafa and Irene Valenzuela}
\end{center}

\vspace{0.15cm}
\begin{center}        
 \emph{Jefferson Physical Laboratory, Harvard University, 
  Cambridge, MA 02138, USA }
 \\[.3cm]

\end{center}

\vspace{2cm}


\begin{abstract}
\noindent

A number of Swampland conjectures and in particular the Trans-Planckian Censorship Conjecture (TCC) suggest that de Sitter space is highly unstable if it exists at all.  In this paper
we construct effective theories of scalars rolling on potentials which are dual to a chain of short-lived dS spaces decaying from one to the next through a cascade of non-perturbative nucleation of bubbles.  We find constraints on the effective potential resulting from various swampland criteria, including TCC, Weak Gravity Conjecture and Distance Conjecture.  Surprisingly we find that TCC essentially incorporates all the other ones, and leads to a subclass of possible dual effective potentials.  These results marginally rule out emergence of eternal inflation in the dual effective theory.  We discuss some cosmological implications of our observations.

\end{abstract}

\thispagestyle{empty}
\clearpage

\setcounter{page}{1}


\newpage

  \tableofcontents

\newpage

\section{Introduction}
One of the major challenges facing present-day cosmology is understanding the nature of the observed dark energy.  The simplest model is to assume that the dark energy is the energy of the minimum energy state of a theory.  An example of this is represented by scalar fields with a potential.  In such a scenario the minima of such scalar potential, if such points exist, would be (meta)-stable solutions to dark energy, leading to de Sitter spaces which seem to be a good approximation to the cosmological observations.
Whether such a scenario would be absolutely stable or only metastable would depend on whether there are lower values of energy at other points in field space.

Such a simple picture seems to be difficult or impossible to obtain in string theory, which has led recently to several swampland conjectures quantifying this difficulty.  The dS swampland conjecture \cite{obied2018sitter} states that the slope of the potential $|V'|/V$ cannot be too small.  Its refinement \cite{Andriot:2018wzk,Garg:2018reu,Ooguri:2018wrx} states that this can only be violated in unstable dS spaces where $V''<0$ and is sufficiently large (compared to $V$).  These conjectures would forbid metastable dS spaces to exist.  Another swampland conjecture, Trans-Planckian Censorship Conjecture (TCC) \cite{bedroya2020trans} which broadly leads to the dS swampland conjecture (with more specific bounds for the slope of the potential), is less restrictive, and in particular does allow for the existence of metastable dS spaces, as long as their lifetime is short.  
The short-lived dS spaces decay by transitioning to a state with lower energy.  In this paper, we study the consequences of such short-lived dS spaces.   In particular, we consider a sequence of transitions from one metastable dS space to the next, nucleated by membranes, and capture this in terms of a dual effective theory of a scalar whose rolling in discrete steps captures these transitions.  This scenario is reminiscent of the inflationary models in \cite{Freese:2004vs,Freese:2006fk}, which also involve a cascade of metastable dS vacua. 

The transitions between nearby dS vacua are severely restricted by swampland conditions.  In particular, TCC puts a strong upper bound on the lifetime of such a transition.  Additionally, we can ask how do other swampland conjectures such as the Weak Gravity Conjecture (WGC) restrict the possibilities.  Indeed WGC leads to the statement that the tension of the membranes which nucleate the decay cannot be too large.  Surprisingly, we find that the fast decay implied by TCC already implies this as a consequence.  Moreover, TCC leads to light enough membranes which in some limits can be viewed as localized excitations.  For sufficiently small cosmological constant the generalized distance conjecture leads to predictions of the mass of the tower of such light states.  We find that the TCC is again compatible with this prediction.  This interwoven relationship between different Swampland conjectures which is also seen in many other contexts is indeed reassuring.

One could ask whether the resulting dual effective potentials that emerge are of the generic type allowed by TCC or the fact that they are generated by dS transitions makes them more restrictive.  Indeed we find that they are more restrictive.  In particular eternal inflation which naively is compatible with TCC is marginally ruled out as being dual to such transitions.  This points to the possibility that eternal inflation is never allowed and is in the swampland as has been suggested in \cite{rudelius2019conditions}.

The organization of this paper is as follows:  In section 2 we review the membrane dynamics which lead to decays of the dS space.  We also derive effective dual potentials capturing such transitions.  In section 3 we apply WGC and TCC to the membrane dynamics.  In section 4 we study the emergent potential and study its properties, and in particular, observe that eternal inflation is not compatible in this dual formulation.   In section 5 we discuss the cosmological implications of our observations.
In section 6 we end with some conclusions.   Some of the technical aspects are presented in the appendices.

\label{sec:intro}

\section{Membrane nucleation in metastable de Sitter}\label{sec:CdL}
The basic point of this paper is to study Swampland constraints in a de Sitter space whose cosmological constant changes via non-perturbative membrane nucleation processes. So we first need to understand how this process takes place, and how it translates to an ``effective potential''. We do both things in this section, relegating most details to the appendices.

\subsection{Review: Thin-wall membrane nucleation}\label{sec:rev}
Let us assume the existence of some metastable de Sitter vacuum. This vacuum should eventually decay to some lower energy configuration. The most standard decay channel is via Coleman-de-Luccia bubble nucleation\cite{Coleman:1977py,Coleman:1980aw}, in which a bubble of true vacuum nucleates inside the false vacuum and starts expanding in an accelerated fashion, almost at the speed of light. This is a non-perturbative semiclassical instability whose transition rate can be estimated in terms of a Euclidean instanton solution,
\beq
\label{Gamma0}
\Gamma = P\, e^{-S}
\eeq
where $S$ is the euclidean classical instanton action and $P$ is some prefactor involving the quantum fluctuations. For the bounce solution to exist, the bubble needs to nucleate with a critical radius $R$ such that the cost of energy of expanding the bubble (the surface tension) is smaller than the energy gain associated with the difference of energies outside and inside the bubble. The result for $S$ and $P$ can be computed in the thin wall approximation, which neglects the physical width of the domain wall in comparison to its critical radius. This is done in appendix \ref{app:CdL}, while here we will only present the results when gravitational corrections are negligible\footnote{Gravitational corrections are negligible when the tension os the bubble $T$ is much smaller than the Hubble scale $\sqrt{\Lambda}$ in Planck units. This approximation will be sufficient for this paper, as we will see that larger values of $T$ are not consistent with the swampland constraints.}.

The critical radius $R$ of the bubble in de Sitter is given by 
\beq
\label{R}
(RH)^2\simeq \frac{1}{1+(R_0H)^{-2}} \ , \quad R_0=\frac{T}{\Delta\Lambda}
\eeq
where $H=\Lambda^{1/2}$ is the Hubble scale and throughout the paper, we will be working in Planck units.  Here $T$ is the tension of the domain wall and  $\Delta \Lambda$ is the difference of vacuum energies on the two sides of the bubble.  Note that $R$
is smaller than the Hubble length, $R\leq H^{-1}$, and that in the flat space limit where $H\rightarrow 0$ we get $R=R_0$.
The instanton action in \eqref{Gamma0} is given by
\beq
\label{S}
S\simeq \frac{ T}{  H^{3}}w(R_0H)\ ,\quad \frac{w(q)}{2 \pi ^2}=\frac{1+2/q^2}{\sqrt{1+1/q^2}}-\frac{2}{q}
\eeq
while the instanton prefactor, up to order one factors, reads
\beq
\label{P}
P\simeq T^2 R^2\simeq \frac{T^2R_0^2}{1+(R_0H)^{2}}
\eeq
More details of the computation of the prefactor can be found in \cite{Garriga:1993fh}. Due to the gravitational effects, it is also possible to have up-tunneling in de Sitter space, but it is much more suppressed if $\Delta\Lambda < \Lambda$ (see appendix \ref{app:CdL}).

In the flat space limit, i.e. when the critical radius of the bubble is much smaller than the Hubble scale, the instanton action and prefactor can be approximated by
\beq
\label{flat}
S\simeq \frac{2\pi^2 T^4}{\Delta \Lambda^3} \ ,\quad P\simeq \frac{T^4}{\Delta \Lambda^2}
\eeq
while $R\simeq R_0$.

Our analysis will be mostly in the thin wall approximation, which we just described. In the opposite limit, when the membrane becomes very thick, there is a decay channel known as the Hawking-Moss transition \cite{hawking1982supercooled}, which dominates over the thin wall Coleman-de-Luccia bubble nucleation. While we focus on the thin wall approximation, we can also put some constraints on the Hawking-Moss scenario, which we describe in appendix \ref{app:HM}.

We finish the review with a couple of comments. In subsequent sections, we study a sequence of successive mild tunnelings that could be effectively described by a smoothly evolving scalar field with a potential.  This can be a good approximation only if we assume that the physical observables do not drastically change from one vacuum to the next.   Because of that, in the following, we focus on cases where the de Sitter minimum decays to a less energetic nearby local de Sitter minimum with positive energy.  This, in particular, implies that
\begin{equation}\label{Betaone}
\Delta\Lambda < \Lambda.\end{equation}
Depending on the model, this process could be repeated multiple times, going through different metastable dS vacua until reaching either an AdS supersymmetric vacuum or decaying to nothing\footnote{It has been shown in certain setups of AdS flux vacua \cite{BlancoPillado:2010et,BlancoPillado:2010df,Brown:2011gt} that there is an alternate decay channel where all of the flux is eaten up all at once and spacetime just ends at a ``bubble of nothing''.  It would be interesting to study if the bubble of nothing in dS can also be understood as a limiting process of the thin-wall transitions we are describing here,  and whether this can be used to put an interesting upper bound on the decay rate of a de Sitter vacuum. }. 

In both cases, we expect a drastic change of the physical observables, either by suffering a Big Crunch or because the vacuum annihilates to nothing. In fact, a drastic change when $\Delta\Lambda$ becomes of order $\Lambda$ is also motivated by a generalization of the swampland distance conjecture applied to the space of metric configurations \cite{lust2019ads} since the flat space limit $\Lambda\rightarrow 0$ is at infinite distance in this field space. 
Therefore, we will not discuss these final transitions here but focus on the chain of CdL transitions that will discharge the positive vacuum energy little by little, but staying on a quasi-de Sitter phase and assuming that the physics does not significantly change in the process.
 
Let us finally remark that in the following we will use the above Coleman De Luccia formulae even if the action \eqref{S} is of order one and there is no exponential suppression. This is justified because the coupling of the domain wall is small, as argued in detail in appendix \ref{app:CdL}.

 \subsection{The effective potential}\label{TEP}
 We have just discussed the dynamics of a universe in which bubbles nucleate and expand in an accelerated fashion. But what precisely does a single observer see, on average? Sitting at the center of her very own static patch, things will not change much and will look approximately de Sitter, until she is hit by a bubble, which nucleated somewhere else. 

After the bubble hits, the vacuum energy has changed by a little bit. Averaging over many transitions, we can replace these discrete jumps in the value of the cosmological constant by an effective scalar $\phi$ with a potential $V(\phi)$.  The characteristics of this potential are in turn determined solely by the fundamental parameters of the membrane picture, $T$ and $\Delta\Lambda$. 

 This allows us to connect directly with the usual quintessence/slow-roll inflation literature, and indeed, over large distances and times the two descriptions are interchangeable\footnote{There are two main differences with the standard picture: at short enough times or length scales, the changes in the vacuum energy are discrete as we just discussed; and as we will see later on, we cannot get just any $V(\phi)$ from the membrane perspective; the potential gets additional constraints. 
}.  

A detailed derivation of the potential can be found in appendix \ref{app:potential}. The basic idea is that to compute the vacuum energy one only needs to compute how many bubbles reach the observer per unit time. At first, it would seem one needs to integrate the bubble production rate over the past lightcone of the observer.  
However, a bubble will not reach the observer if it hits another bubble and annihilates with it first. As a result, we only need to integrate the bubble production rate over some spherical effective volume $\mathcal{V}_{\text{eff}}$. Thus, the number of bubbles per unit of proper time is
\begin{equation} \frac{dN}{dt}= \Gamma\, \mathcal{V}_{\text{eff}}.\label{mfp0}\end{equation}
Equation \eqref{mfp0} is all we needed to compute the potential explicitly since
\begin{equation}\frac{dV}{dt}=\Delta\Lambda \frac{dN}{dt}=\Delta\Lambda \Gamma\, \mathcal{V}_{\text{eff}}\sim \frac{(V')^2}{\sqrt{V}},\end{equation}
where the last equality is a slow-roll expression. That is, we assumed that the vacuum energy can be described by a slow-rolling scalar with potential $V(\phi)$, and then equated the slow-roll expression for $dV/dt$ from what we get from membranes. Rearranging, one gets
\begin{equation}
\label{V}
\left(\frac{V'}{V}\right)^2\sim \frac{\Delta\Lambda}{\Lambda^{3/2}} \Gamma\, \mathcal{V}_{\text{eff}},\end{equation}
which determines the potential completely once we know $\mathcal{V}_{\text{eff}}$. A detailed derivation of this effective volume can be found in appendix \ref{app:potential} (see \eqref{Veff} for the general result for $\mathcal{V}_{\text{eff}}$.). Here, we will only note the two limiting cases that are relevant for our constraints:\begin{itemize}
\item It is intuitively obvious that $\mathcal{V}_{\text{eff}}$ cannot grow larger than the Hubble horizon. In case that $\mathcal{V}_{\text{eff}}$ is this large, we find
\begin{equation}\frac{V'}{V}=\frac{\Delta\Lambda^{1/2}}{\Lambda^{3/2}}\, \Gamma^{1/2}.\label{pot11}\end{equation}
 This corresponds to the case where collisions are rare ($\Gamma\ll H^4$) and basically all membranes which are produced in the past lightcone reach the observer.
\item On the other hand, when the critical radius is much smaller than the Hubble scale and collisions are common ($\Gamma\gg H^4$), the effective volume is determined by the distance to the closest nucleating event, which is of order $\Gamma^{-1/4}$.  So in this case
\begin{equation}
\label{VV}\frac{V'}{V}=\frac{\Delta\Lambda^{1/2}}{\Lambda^{3/4}}\, \Gamma^{1/8}.\end{equation}
\end{itemize}

So to sum up, we have membranes that discharge the background cosmological constant, and a convenient description in terms of an effective potential. Without any further assumptions, this could take a very long time, the effective potential would be extremely flat, and the de Sitter could be extremely long-lived. In this paper, we will see that Swampland conditions such as TCC and WGC place constraints on just how fast these decays can happen.

\section{Swampland constraints on bubble nucleation}

The goal of this section is to investigate the swampland constraints on the decay rate of bubble nucleation in metastable de Sitter vacua.  We will see that, in particular, the Weak Gravity Conjecture and the Transplanckian Censorship Conjecture imply non-trivial constraints on the properties of the bubbles/membranes. It will be very convenient from now on to parametrize the scaling of the tension of the bubble and the difference in the vacuum energy in terms of $\Lambda$ as follows,
\beq
\label{ab}
T\sim \Lambda^\alpha \ ,\quad \Delta\Lambda \sim \Lambda^\beta.
\eeq
For the time being, we can think of $\alpha,\beta$ as constants although we will later allow them to depend on $\Lambda$ as well.

\subsection{The Weak Gravity Conjecture}\label{WGC}

The Weak Gravity Conjecture \cite{ArkaniHamed:2006dz} states that, given a theory with a p-form gauge field weakly coupled to Einstein gravity, there must exist an electrically charged state satisfying
\beq
\label{WGC0}
 \gamma \, T\leq Q
\eeq
where $Q=g_pq$ is the physical charge (including the gauge coupling $g_p$), $T$ is the tension and $\gamma$ is the charge to tension ratio of an extremal black brane in that theory. We will be applying this to a codimension-1 object, a membrane coupled to a 3-form gauge field with gauge coupling $g_3$.  Since our primary interest is weakly curved de Sitter space, we will be ignoring potential corrections to the WGC bound from the positive cosmological constant\footnote{See \cite{Huang:2006hc,Montero:2019ekk,Antoniadis:2020xso} for work on the (particle) version of the WGC in dS.}. 

The interpretation of the WGC for codimension-one objects is a bit subtle as the backreaction of these objects is very strong and destroys the asymptotic structure of the vacuum. Hence, they should not be understood as normalizable states around a given vacuum but rather as defects sourcing localized EFT operators \cite{Michel:2014lva}---a perspective that has been recently analyzed in \cite{Lanza:2020qmt} in relation to the Swampland conjectures---. The 4d backreaction away from the object translates then into a classical RG flow to low energies that makes the tension scale-dependent. In this paper, we will assume that the WGC applies to any energy scale and will impose the WGC to the domain wall solutions in the IR (an approach already taken in \cite{Ooguri:2016pdq} when using the WGC to argue for a bubble instability  for any non-supersymmetric vacuum).

In order to apply the WGC to the domain walls, we are assuming that the CdL bubble nucleation corresponds to a Brown-Teitelboim transition \cite{Brown:1988kg} in the sense that the domain wall contains a localised membrane on its core charged under a 3-form gauge field. This is characteristic for example of vacua arising from compactifications with internal fluxes. When crossing one WGC domain wall of quantized charge $q$, the quantized background 4-form field strength $F_4$ changes by $q$ units. Since this field strength parametrises the vacuum energy, the charge of the domain wall roughly corresponds to the difference of vacuum energies in the tunneling transition. Consider for instance a single 3-form, with a (possibly field-dependent) gauge coupling $g_3$. The vacuum energy is  such that the potential reads

\begin{equation}\Lambda=\frac12 g_3^2n^2.\label{gn}\end{equation}

We allow $g_3$ to depend on $n$ polynomially, and assume that any other contribution to the vacuum energy is subleading with respect to \eqref{gn}. Then, one has that
\begin{equation}
\label{Q2}
 Q= g_3\,\Delta n\simeq  \Delta (\sqrt{\Lambda}).
 \end{equation}

By plugging \eqref{Q2} into \eqref{WGC0} we get that the WGC for domain walls implies
\begin{equation}
\label{TQ} T\lesssim \frac{\Delta \Lambda}{\Lambda^{1/2}}\end{equation}
where we have assumed that the variation of vacuum energy is
small. We have also neglected an order one factor coming from the extremality factor $\gamma$ in \eqref{WGC0} as we will only be interested in the scaling of the tension with the vacuum energy. Upon using \eqref{ab}, the above inequality translates into the following constraint on $\alpha,\beta$,
\beq
\alpha-\beta+\frac12\geq 0 \ .
\label{WGCab}
\eeq
It is interesting to note that \eqref{TQ} is equivalent to imposing $R_0\lesssim H^{-1}$ where $R_0$ is the flat space radius defined in \eqref{R}. Recall from section \ref{sec:CdL} that the decay rate can be written as a function of two variables, $T$ and $R_0$ in Hubble units. The WGC bound $R_0\lesssim H^{-1}$ makes the instanton action small, so ameliorates the exponential suppression, but also decreases the prefactor, as can be checked using \eqref{S} and \eqref{P}. Hence, for a given value of the tension, the WGC implies an upper bound on the decay rate of bubble nucleation. This is reminiscent of the situation in \cite{Montero:2019ekk}, where WGC-like considerations led to an upper bound on how fast black holes should decay.

\subsection{Transplanckian censorship conjecture}\label{TCCsec}

dS space seems to be difficult to realize in controllable regimes of String Theory. An example of this tension is a class of no-go theorems that forbid a metastable dS in the asymptotic of the field space which motivated the dS Swampland conjecture \cite{obied2018sitter} (for related Swampland ideas see \cite{Andriot:2018wzk,Garg:2018reu,Ooguri:2018wrx,Danielsson:2018ztv,Grimm:2019ixq,Andriot:2020lea,Danielsson:2018qpa,Hebecker:2018vxz,Klaewer:2018yxi,Andriot:2018mav,lust2019ads,Montero:2019ekk,Lanza:2020qmt} ). This key observation has led to multiple Swampland conditions that aim to find a more general principle that could explain the tension between the dS space and consistent quantum theories of gravity. One of such Swampland conditions, the Trans-Planckian Censorship Conjecture (TCC), states that the expansion of the universe must slow down before all Planckian modes are stretched beyond the Hubble radius  \cite{bedroya2019trans}. If TCC gets violated, the Planckian quantum fluctuations exit the Hubble horizon, freeze out and classicalize which is, at the very least, strange. A variety of non-trivial consequences of TCC for scalar field potentials were studied in  \cite{bedroya2019trans} and shown to be consistent with all known controllable string theory constructions.  In this paper, we will not enter into motivating the TCC, but simply study its implications for the case of metastable de Sitter vacua in more detail. A survey of the motivations for TCC can be found in \cite{bedroya2020}.

For metastable de Sitter spaces where the Hubble parameter stays constant, the TCC imposes an upper bound on the lifetime  as follows \cite{bedroya2019trans},
\beq
\tau\lesssim \frac{1}{H}\frac{1}{\log(1/H)}.
\eeq
 We will be referring to this upper bound as the TCC time $\tau_{TCC}$. 
  In the rest of this paper, we focus on the leading terms in our computations and will ignore the logarithmic correction factor above. We will come back and discuss the effect of the subleading corrections in subsection \ref{sec:o1}.  
 
Let us now study the consistency of the TCC with the CdL decay mechanism reviewed in section \ref{sec:CdL}. In particular, we will momentarily see that thin-wall tunneling could be consistent depending on the characteristics of the domain wall. In appendix \ref{app:HM} we also discuss the Hawking-Moss transition to show that when it is the dominant decay channel it is (marginally) inconsistent with the TCC.

In section \ref{sec:rev} we provided $\Gamma=Pe^{-S}$ in terms of $T$ and $\Delta\Lambda$ for thin-walls. By plugging \eqref{ab} into \eqref{S} and \eqref{P}, we find the TCC takes the following form in terms of $\alpha$ and $\beta$,
\beq\label{TCC}
\Gamma > H^4\ \rightarrow \ \frac{\Lambda^{4\alpha-2\beta -2}}{1+\Lambda^{2\alpha-2\beta+1}}\exp(-\Lambda^{\alpha-3/2}w(\Lambda^{\alpha-\beta+1/2}))\gtrsim 1.
\eeq
When $\Lambda$ is very small, the above inequality can be approximated by
\beq\label{TCC2}
\Lambda^{4\alpha-2\beta -2}\exp(-\Lambda^{4\alpha-3\beta})\gtrsim 1,
\eeq
which can also be derived by using the flat space approximations for $P$ and $S$ in \eqref{flat}.

\subsection{Constraints on domain walls}\label{3}
In the previous two subsections, we discussed how the Swampland conditions could be applied to the domain walls. In this subsection, we combine those results and perform a systematic study of what domain walls belong to the Swampland. 

\begin{figure}[!htb]
\begin{center}

\includegraphics[width=0.55\textwidth]{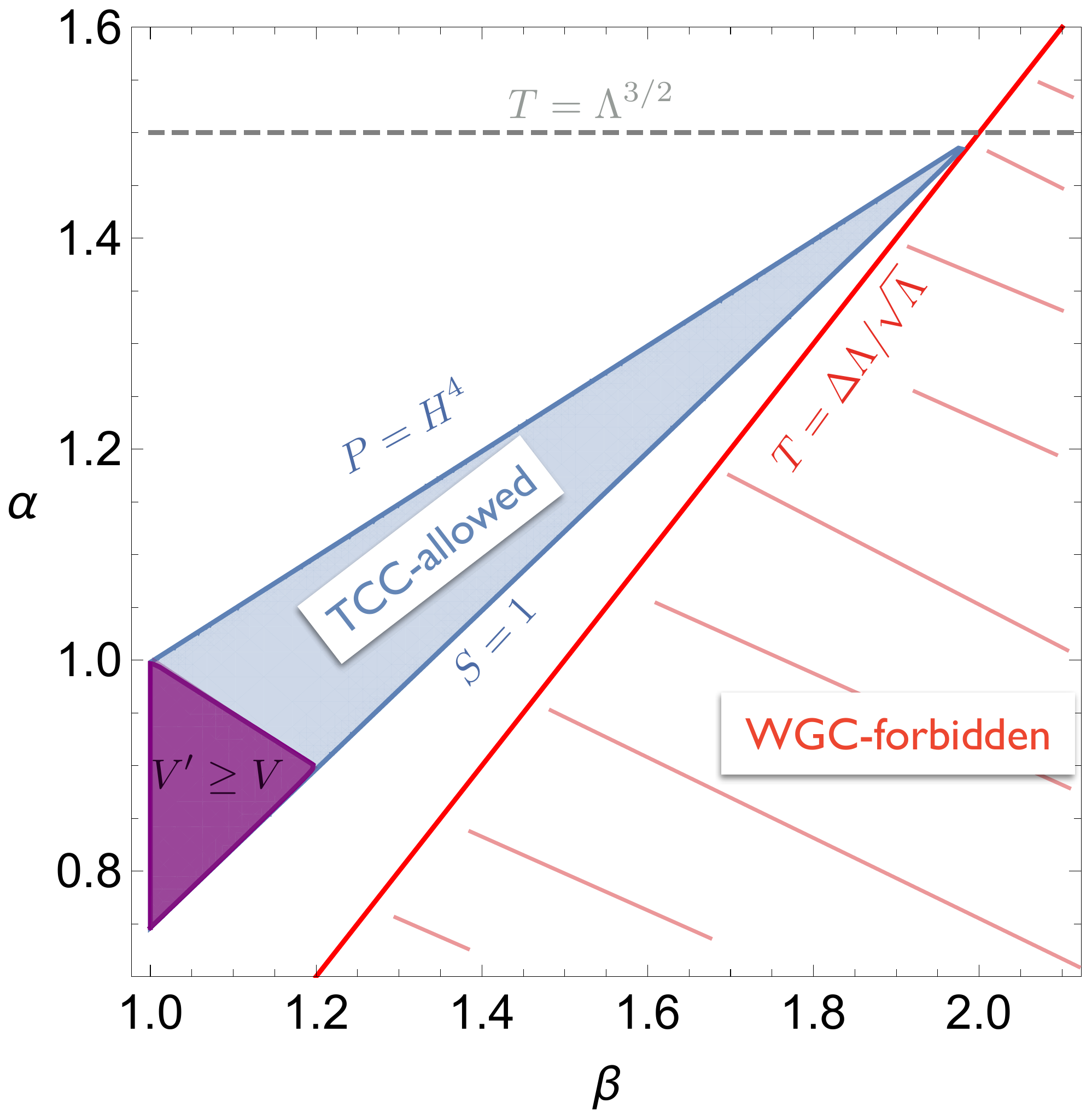}

\end{center}
\vspace{-0.7cm}
\caption{Allowed regions in the $(\alpha,\beta)$ plane for $\Lambda=10^{-120}$.  The left hand side of the red line is allowed by the WGC for membranes \eqref{WGCab}, while the light blue region corresponds to the TCC allowed region \eqref{TCC}. The purple region corresponds to $V'/V>1$ and points above the horizontal dotted line have $T\leq \Lambda^{3/2}$, which are disfavoured by the Higuchi bound and Distance Conjecture.}\label{regions}
\end{figure}

Figure \ref{regions} shows how the WGC (eq. \eqref{WGCab}) and the TCC (eq. \eqref{TCC}) constrain the values of $\alpha$ and $\beta$ which characterize thin-walls to lie in a confined blue region. We only present the results $\beta>1$ as this is implied by \eqref{Betaone}, taking into account that in Planck units $\Lambda <1$.  An interesting feature is that TCC imposes a stronger constraint than WGC; in other words, in most of the parameter space, TCC implies WGC for domain walls. There is a region near $(2,3/2)$ where the two curves intersect. Which one imposes the stronger constraint is sensitive to $\mathcal{O}(1)$ factors; we will comment on these in section \ref{sec:o1}.

The boundary of the blue region, which represents the TCC condition \eqref{TCC}, does not have a simple analytic form in $\alpha$ and $\beta$ for a general $\Lambda$.  However, for exponentially small values of $\Lambda$, such as would be in our universe, the blue region given by $\Gamma=Pe^{-S}\gtrsim H^4$ can be well approximated by a triangle whose boundaries can be easily determined by looking at \eqref{TCC2}, from which we can extract $P\simeq \Lambda^{4\alpha-2\beta}$ and $S\simeq \Lambda^{4\alpha-3\beta}$. The triangle is delimited by two lines, one corresponding to $P\gtrsim H^4$, and another to $S\lesssim\mathcal{O}(1)$ to eliminate the exponential suppression, as follows
\begin{equation}S\leq 1\ \rightarrow \ 4\alpha\geq 3\beta,\label{tcc1}\end{equation}
\begin{equation}\mathcal{P}\geq H^4\ \rightarrow\ 4\alpha-2\beta\leq 2\label{tcc2}\end{equation}
These two lines provide a fairly accurate envelope of the numerical blue region if $\Lambda$ is very small, except for the region at the tip of the triangle. The point where the triangle almost touches the WGC line corresponds to eternal inflation potentials, as we will discuss in more detail in section \ref{sec:eternal}.

We also note in passing that for the whole approach to be valid, we should impose that the radius of the bubbles is above the cutoff of the EFT. Choosing the cutoff to be at the Planck scale, this just removes the point $(\alpha,\beta)=(1,1)$, as all the bubbles inside the blue region have a subplanckian radius. Lowering the cutoff from the Planck scale to e.g. GUT scale would remove a very small region around this point, but this does not affect our constraints very much and the qualitative features of the plot remain unaltered. 

We have also added a few more lines in figure \ref{regions}. First, the horizontal dotted grey line represents membranes with $T= \Lambda^{3/2}$. In the next section, we will provide some arguments that motivate us to only allow membranes below this line.
Finally, we have highlighted in purple the region corresponding to $V'/V>1$ by using the full derivation of the effective potential in terms of the decay rate in \eqref{V}. This region is excluded by observational constraints in our universe. The line bounding this region and the rest of the blue triangle can be simply derived from  \eqref{VV}, which is a good approximation since $\Gamma>H^4$ inside the blue region. Hence, by plugging $\Gamma \simeq P\simeq \Lambda^{4\alpha-2\beta}$ into \eqref{VV} we get
\beq
\frac{V'}{V}=1 \ \rightarrow\  2\alpha+\frac{\beta}{4}-\frac34=0\ .
\eeq

\subsection{Comments on Higuchi bound and Distance Conjecture } \label{app:WGC}
In figure \ref{regions} we have drawn a horizontal dotted line at the value associated with $T=\Lambda^{3/2}$.  Here we will provide three different arguments in favor of imposing $\alpha \leq 3/2$ which, even if not completely conclusive, motivates this upper bound.  These arguments involve (1) the breakdown of effective field theory (2)  applying Higuchi bound to the membrane and (3) the application of membrane excitations as leading to light states predicted by the generalized distance conjecture  \cite{lust2019ads}.  Note that the combination of this upper bound with the WGC constraints implies a finite region on the $(\alpha,\beta)$-plane, which implies by itself an upper bound on the decay rate (independently of the TCC). Interestingly, this upper bound is a bit less restrictive but still consistent with the TCC.

Our domain wall solutions contain fundamental membranes on their core, which mediate transitions between different flux vacua. Often in string compactifications, the domain wall solutions involve additional scalar fields which get a nontrivial profile in the membrane background. If we go high up in energies, the membranes can be seen as localized free objects with a tension $T_{\rm mem}$  which can differ from the tension of the domain wall due to the contribution from the scalar flow driven by the membrane backreaction, so that $T\geq T_{mem}$. For the semiclassical description of these membranes not to break down, we need the tension to be above the cut-off of the EFT, $T_{\rm mem}\geq \Lambda^{3/2}_{\rm cutoff}$ \cite{Lanza:2020qmt}. Otherwise, it would not be possible to describe the membrane within a local EFT. Since this cut-off is associated with the membrane sector, it can be disconnected from the SM of particle physics and could, in principle, take any value. However, it seems reasonable to impose that it is above the Hubble scale in an expanding universe, $\Lambda_{\rm cutoff}>H$.

Hence, we get that $T_{\rm mem}\geq \Lambda^{3/2}$ so there is a lower bound for the membrane tension in terms of the cosmological constant, which in turn implies a lower bound for the domain wall tension in the IR as  $T\geq T_{mem}\geq \Lambda^{3/2}$, implying
\beq
\label{alpha}
\alpha\leq 3/2\ .
\eeq
This is also consistent with the fact that in dS space, any mass scale less than Hubble is physically not measurable in the current phase of the universe.  In other words, having the mass scale associated with the membrane $T^{1/3}\leq \Lambda^{1/2}$ will be unobservable.  So we may as well restrict to $\alpha \leq 3/2$.
To sum up, as long as the domain wall has a fundamental membrane on its core which can be described semiclassically with a local EFT, and the EFT cut-off is bigger than the Hubble scale, then one needs to impose \eqref{alpha}.  Of course, not every CdL transition needs to have the interpretation of a Brown-Teitelboim flux transition with a fundamental semiclassical membrane on its core, but otherwise, the justification of applying the WGC to the domain walls is less clear. If the membrane cannot be described semiclassically, then the EFT is non-local and it is not clear how to even start defining a charge under a gauge field and how to apply the WGC then.

The second argument comes from applying the Higuchi bound to the membranes.  First of all, notice that it is not possible to apply the Higuchi bound directly to the IR domain walls, as the membranes are confined and the mass scale of the excitation modes is not associated with $T^{1/3}$. In fact, if the bubble is expanding, the volume contribution in $E\sim TR^{2}-\Delta\Lambda R^3$ dominates over the tension surface, and the modes are tachyonic as they are describing a vacuum instability. 
However, it is possible to apply the Higuchi bound to the localized membranes at the core of the domain walls as long as the relevant energy scale is above the confinement scale and they behave as free objects.
Indeed, the condition $T_{\rm mem}\geq \Lambda_{\rm cut-off}^{3/2}$ guarantees that there is a regime in energies in which very small spherical membranes behave as semiclassical unstable particles with a lengthscale at least of the order of their Compton length $l_c\sim T_{\rm mem}^{-1/3}$. 

 In other words, there are small unstable pockets that contract as soon as they are formed, with an energy that it is well approximated by $E\sim T_{\rm mem}l_c^2\sim T^{1/3}_{\rm mem}$  since $T_{\rm mem}l_c^2 \gg \Delta\Lambda\, l_c^3$ if  $T_{\rm mem}\geq \Lambda^3_{\rm cutoff}$ and $\Delta\Lambda \ll  \Lambda_{\rm cutoff}$\footnote{Notice that $T_{\rm mem}l_c^3 \gg \Delta\Lambda\, l_c^3$ is equivalent to require $S\sim T_{\rm mem}^4/\Delta\Lambda^3 \gg 1$. Thanks precisely to the scalar contribution due to the backreaction induced by the membrane, we can satisfy this condition for the membranes but still violate it for the domain wall in the IR (so that the domain wall will be consistent with the TCC later on). For this to happen one needs to have $T/Q|_{DW} < (T/Q)_{\rm mem}$, which is expected by the WGC if we have a vacuum which breaks spontaneously supersymmetry but the membranes were originally BPS.}. By applying the Higuchi bound to these small spherical membranes, we get that $T^{1/3}_{\rm mem}\geq \Lambda^{1/2}$ implying again \eqref{alpha}. 

The last argument is more a proposal for an interpretation of the role of these membranes in case they satisfy \eqref{alpha}.
Interestingly, these membranes are candidates to fulfill the AdS Distance Conjecture in de Sitter space \cite{lust2019ads}. The conjecture states that there should an infinite tower of states with a mass of order
\beq
m\sim \Lambda^\delta
\eeq
since the flat space limit $\Lambda\rightarrow 0$ is at infinite distance in the space of metric deformations. In de Sitter space, the Higuchi bound forces $\delta$ to be $\delta \leq 1/2$, which is equivalent to having $T_{\rm mem}\geq \Lambda^{3/2}$.  The conjecture does not specify what is the origin of the tower of states. In AdS space, they usually correspond to particles, KK towers for concreteness, underlying the absence of scale separation typically observed in AdS vacua. 
An interesting possibility is that, in de Sitter space, the tower of states comes from membranes and it is, therefore, eventually underlying the instability of these vacua. We could turn the argument around and say that, if the membranes provide the states satisfying the AdS Distance Conjecture, then they need to satisfy \eqref{alpha}.

\section{Emergent Potential and the Swampland}

In the previous section, we studied how the Swampland conditions constrain the domain wall parameter space. As we saw in subsection \ref{TEP}, successive tunnelings between neighboring vacua can be effectively described by a smooth rolling of an emergent scalar field in a potential. We can either apply TCC in the membrane perspective or directly to the emergent effective potential without taking its microscopic origin into account. We will find that the membrane perspective leads to a restrictive class of emergent potentials that could not be obtained otherwise.  This seems to extend the meaning of TCC and in particular, leads to essentially forbidding eternal inflation. 

\subsection{Flat potentials and TCC}\label{MVQ}

The TCC implies the general statement that a quasi-deSitter phase cannot last more than ${1\over H} \ln(1/H)$. We will now tailor this statement to the particular case of very flat ($|V'|\ll \frac{V}{|\ln(V)|}$) monotonic potentials. As we will see in subsection 4.2, these are the kind of potentials we get from the membrane picture on a range of parameter spaces. We aim to find the strongest condition that TCC alone imposes on this kind of potential.

First, we show that the TCC implies the field range needs to be sub-Planckian. We prove this by contradiction. Suppose the field range is trans-Planckian. Consider a slow-roll trajectory over an $\mathcal{O}(1)$ sub-interval of the field range. For the slow-roll trajectory we have
\begin{align}\label{srtr}
    d\phi\simeq \frac{|V'|}{\sqrt{3V}}dt.
\end{align}
 Since $|V'|\ll V$, the change in $V$ over this field range is negligible and $V$ can be taken to be constant. From the TCC, we find $\Delta t< \frac{1}{H}\ln(\frac{1}{H})\sim \frac{|\ln(V)|}{\sqrt{V}}$. Plugging this and $|V'|\ll \frac{V}{|\ln(V)|}$ in equation \eqref{srtr} gives
\begin{align}
    \Delta\phi\sim \frac{|V'\ln(V)|}{V}\ll 1,
\end{align}
which is in contradiction with our assumption. Thus, the field range must be sub-Planckian. This fact combined with the fact that the potential is very flat, allows us to take $V$ and $H$ to be nearly constant in our computations. Given that $H$ is almost constant, the consistency of the slow-roll trajectory with TCC would imply that any other trajectory is consistent with TCC as well.   This is because it only takes one Hubble time for a trajectory to become slow-roll and TCC upper bound for the duration of the inflation is $\frac{1}{H}\ln(\frac{1}{H})$ which for small values of the cosmological constant is much greater than the Hubble time. So the first part of the trajectory before the slow-roll is negligible. In any case, note that the derivation of the effective potential \eqref{V} in appendix \ref{app:potential} assumes slow-roll.

This can also be expressed in terms of the potential alone, without referring to the slow-roll trajectory. By rearranging \eqref{srtr} and imposing TCC, we get
\begin{equation}
         \sqrt{3V}\int\frac{d\phi}{|V'|}=\int dt<\sqrt\frac{3}{V}\ln(\sqrt\frac{3}{V})\quad
       \rightarrow \quad 2V\int \frac{d\phi}{|V'|}\lesssim|\ln(V)|.
\end{equation}
 
In short, TCC only imposes that the potential must get steep ($|V'|\gtrsim V$) after the time $\tau_{TCC}\sim 
\frac{1}{H}\ln(\frac{1}{H})$. When the potential is induced by successive tunnelings as in section \ref{TEP}, this constraint could be interpreted as a statement about the time evolution in the $\alpha-\beta$ plane in figure \ref{regions}. TCC is equivalent to requiring the trajectory in $\alpha-\beta$ plane to reach the purple region ($|V'|\gtrsim V$) in less than $\tau_{TCC}$ time, and nothing else. In particular, it does not lead to any pointwise constraints on the potential. As we will now see, combining with the membrane picture, it is possible to do better.

\subsection{Swampland constraints on the membrane effective potential}

We start by finding the characteristics of the effective potentials that can arise from membrane tunneling. Suppose we have a nearly flat ($|V'|\ll V$) monotonic potential. We investigate the possibility of dividing up the field range into small enough intervals $(\Delta\phi)_i$ such that each piece can be approximated by a linear function, and  
each discrete jump can be realized by an allowed membrane nucleation. We define parameters $\theta_n$ and $\gamma_n$ for the $n$-th piece as follows.
\begin{align}
    &V_n^{\theta_n}=\frac{|V'_n|}{V_n},\nonumber\\
    &V_n^{\gamma_n}=(\Delta\phi)_n,
\end{align}
where $V_n$ and $V'_n$ are the potential and its slope at the n-th interval. Supposing $\Delta\phi$ is small enough we find
\begin{align}
    V_n^{\beta_n}=(\Delta V)_n\simeq|V'_n|(\Delta\phi)_n
\end{align}
which gives the following relation for $\beta$,
\begin{align}\label{betaa}
    \beta_n=\theta_n+\gamma_n+1.
\end{align}
Applying the slow-roll condition gives
\begin{align}
    (\Delta t)_n\simeq\frac{(\Delta V)_n}{|V'_n|\dot\phi_n}\sim\frac{(\Delta V)_n\sqrt{V_n}}{|V'_n|^2}=V_n^{\beta_n-2\theta_n-\frac{3}{2}}.
\end{align}
Plugging $\beta$ in terms of $\theta$ and $\gamma$ leads to 
\begin{align}\label{1234}
    (\Delta t)_n\sim V_n^{\gamma_n-\theta_n}\frac{1}{H}
\end{align}

\begin{figure}[!htb]
\begin{center}
\includegraphics[width=0.6\textwidth]{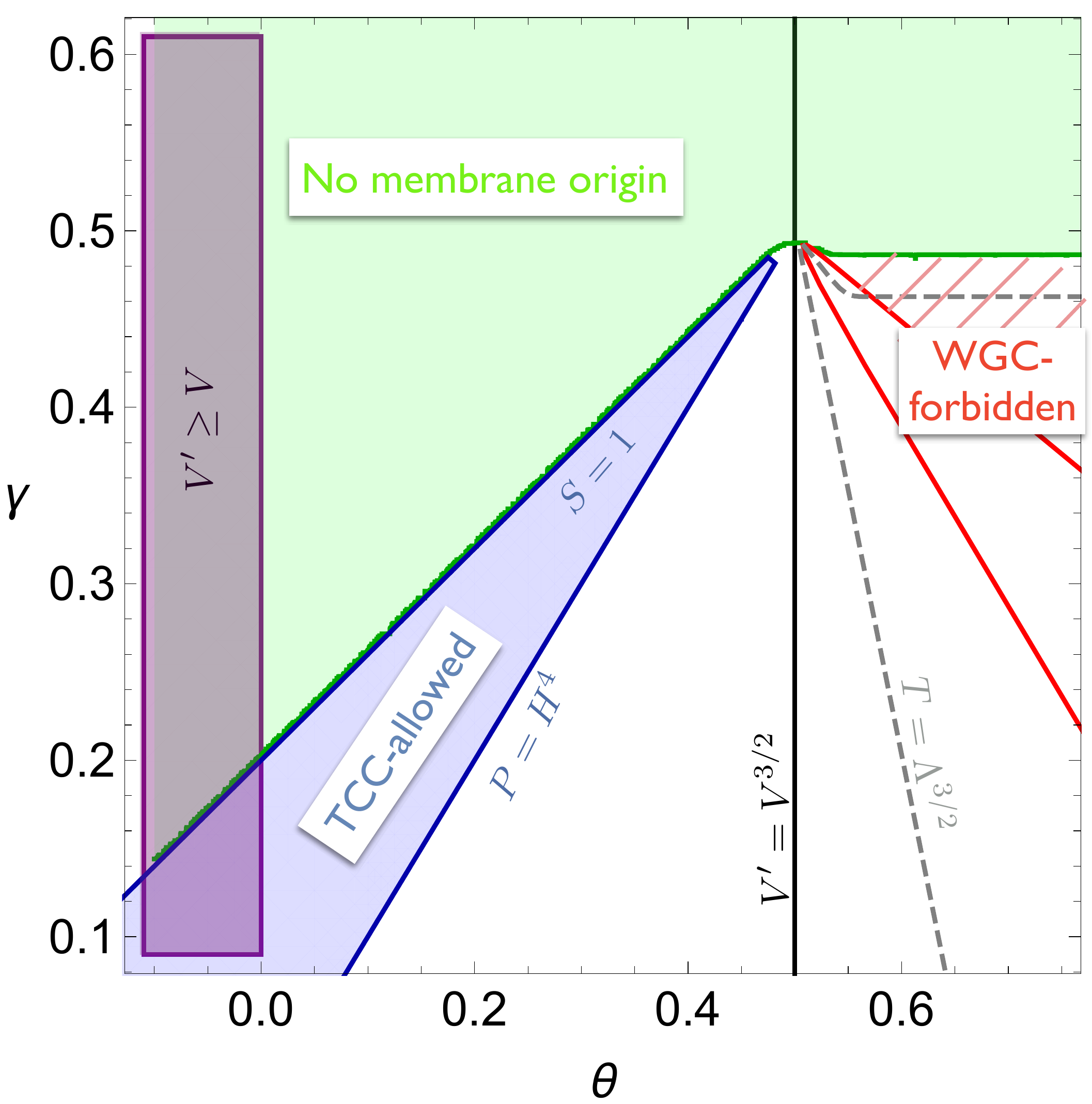}
\end{center}
\vspace{-0.7cm}
\caption{Allowed regions in the $(\gamma,\theta)$ plane for $\Lambda=10^{-120}$.  Not all of the $(\theta,\gamma)$ plane corresponds to a valid membrane picture (i.e. the map from the $(\alpha,\beta)$ plane to $(\theta,\gamma)$ is not onto); we have shaded in light green the region which is not part of the image. The purple region corresponds to $V'/V>1$, while the light blue region corresponds to the TCC allowed region. The red line saturates the WGC for membranes and the region above the upper red branch is forbidden by the WGC. The region to the right of the grey dotted line one has $T< \Lambda^{3/2}$ and is disfavoured by the Higuchi bound and Distance Conjecture. The black line represents the ``eternal inflation'' locus, defined by $|V'|=V^{3/2}$. }
\label{regions-gt}
\end{figure}

Note that the derivation of the effective potential in section \ref{TEP} allows us to compute $\theta$ as a function of $\alpha$ and $\beta$ defined in \eqref{ab}. Using this as well as \eqref{betaa}, we can translate the swampland constraints on the $(\alpha,\beta)$-plane of figure \ref{regions} to the $(\theta,\gamma)$-plane instead, as shown in figure \ref{regions-gt}. It is very important that not every point in the $(\theta,\gamma)$ is the image of a point in the $(\alpha,\beta)$ plane; points that are not in the image (green region in the figure) are not physically meaningful from the point of view of the membranes. In addition, the map is 2-to-1; two different points in the $(\alpha,\beta)$ plane map to the same point on  $(\theta,\gamma)$\footnote{This is related to the fact that the lifetime of the dS can be unchanged if while the membrane action increases the prefactor increases in a way to compensate this.}. The blue region in figure \ref{regions} ``folds over itself'' along $S\sim 1$ to be mapped to the blue region in figure \ref{regions-gt}. Every point in the blue region in figure \ref{regions-gt} has two preimages; one with $S>1$ and the other with $S<1$. 
More generally, the entire $(\alpha,\beta)$ plane  folds over itself  along the curve $\partial_\alpha (\Gamma)=0$, 
 which is very close to, but not exactly at, the lower boundary of the TCC region in figure \ref{regions}, and after the vertex of the TCC triangle it goes on to a line of almost constant $\alpha$. The folding line gets mapped to the boundary of the green region in figure \ref{regions-gt}, which is approximately described by the following function: \begin{equation}\gamma=\begin{cases} 
     \frac15 \left(1+3\, \theta\right) & \theta\lesssim 1/2 \\
     \theta-0.013 & \theta \gtrsim 1/2\end{cases}\label{eo2}.\end{equation}
This provides, for each value of $\theta$, the maximum value of $\gamma$ consistent with a membrane origin of the effective potential. Notice that $\gamma= \frac15 (1+3\, \theta) $ is equivalent to the condition for the instanton action to be $S\sim 1$ in the flat space limit.

There is a potential ambiguity in the definition of the decay time that needs to be addressed. The time $\Delta t$ in \eqref{1234} is the time it takes for the transition to occur everywhere in the Hubble patch which is different from the lifetime associated to an individual bubble\footnote{The lifetime associated to an individual membrane is the time scale for only one bubble to form somewhere in the universe and shift the value of the potential by $\Delta V$ within that bubble. Imposing that this time scale is smaller than Hubble time is equivalent to the TCC constraint for membranes imposed in section \ref{TCCsec}, i.e. $\Gamma> H^4$. The homogeneous time scale $\Delta t$ in \eqref{1234} is when the average of $V$ over the whole Hubble patch decreases by $\Delta V$ and is given by $\Delta t =\Gamma^{-1}\mathcal{V}_{\rm eff}^{-1}$, where the effective volume $\mathcal{V}_{\rm eff}$ is derived in \eqref{vol}.} used when applying the TCC in section \ref{3}. Applying the TCC to the former time, i.e. $\Delta t \leq H^{-1}$ simply implies the condition $\gamma \geq \theta$, which a priori might seem different (even weaker) than the constraint coming from applying the TCC to the membrane picture ($\Gamma > H^4$) represented as the blue region in figure \ref{regions}. However, as we see from the figure, in practice applying the TCC to the effective potential provides the same constraints as applying the TCC to the individual membranes as long as we restrict ourselves only to those potentials that can be interpreted as originating from averaging over a cascade of membrane nucleation transitions\footnote{Using \eqref{vol} one could analytically check the equivalence between $\Delta t=\Gamma^{-1}\mathcal{V}_{\rm eff}^{-1}< H^{-1}$ and $\Gamma> H^4$ by noting that  $\mathcal{V}_{\rm eff}$ takes values in between  $H^{-3}$ (when the decay rate is small) and $\Gamma^{-3/4}<H^{-3}$ (when the decay rate is large).}. This is because the top boundary of the blue region coincides with the limit of the region admitting a membrane origin.

In figure \ref{regions-gt}, there is a maximum value of $\theta$ allowed by TCC\footnote{This feature is sensitive to $\mathcal{O}(1)$ factors, but we will show in the next subsections that the TCC in combination with the WGC always implies an upper bound on $\theta$.
}. In subsection \ref{MVQ} we saw that TCC by itself does not bound $\theta$ in general, so the constraint comes from the assumption that the potential has a fundamental description in terms of membranes. The membrane picture strengthens TCC, turning it into a constraint on the potential. 

Finally, it is important to note that while $\gamma$ is a physical observable since it quantifies inhomogeneities in the bubble nucleation process, this is a piece of information that gets lost in the effective potential description, which only tracks the Hubble scale evolution. In other words, the only constraint that determines whether a potential can be chopped into pieces generated by membranes is the upper bound on $\theta$. 

\subsection{No Eternal Inflation\label{sec:eternal}} 
So what general lessons can we learn from the membrane picture? We will now argue that the eternal inflation point is marginally excluded by our constraints. 

As could be seen in figure \ref{regions-gt}, the maximum allowed values for $\gamma$ and $\theta$ are realized when  the TCC and/or the WGC get saturated and hit the boundary of the no-membrane origin region, so that the lower boundary of the blue region and/or the red line intersect  \eqref{eo2}. The intersection of these three curves nearly happens at the same point which, by using  \eqref{1234}, satisfies
\beq
\gamma_{max}=\theta_{max}+\dots
\eeq
where the ``$\dots$'' denote subleading corrections that go away in the limit $\Lambda\rightarrow 0$.
From equation \eqref{betaa}, we find that $\beta$ is maximized at this point as well,
\begin{align}\label{tbe}
    \theta_{\max}\simeq \frac{\beta_{\max}-1}{2}+\dots
\end{align}

As discussed in subsection \ref{WGC}, applying the WGC to the $\beta_{\max}$ point implies $\alpha\geq\beta-\frac{1}{2}$, while the saturation of TCC implies $P\sim H^4$, where $P$ is the prefactor of the decay rate defined in \eqref{P}  as discussed in section \ref{TCC}. Using \eqref{P} we find
\begin{align}
\label{Pet}
    P\sim H^4\rightarrow \frac{\Lambda^{4\alpha-2\beta}}{1+\Lambda^{2\alpha-2\beta+1}}\sim \Lambda^2.
\end{align}
For $\Lambda$ small, the denominator becomes an order one factor $1+\Lambda^{2\alpha-2\beta+1}\sim \mathcal{O}(1)$. 
Plugging in $\alpha\geq\beta-\frac{1}{2}$ gives
\begin{align}
\label{betamax}
  \Lambda^{2\beta_{\max}-2}\gtrsim\mathcal{O}(1)\times \Lambda^2\ \rightarrow\  \beta_{\max}\lesssim 2+ \dots\end{align}
The sign of the next to leading term above depends on the value of order one factors coming from the prefactor as well as corrections to the TCC and the WGC. We will comment on the effect of these corrections in section \ref{sec:o1}.

Plugging the above inequality in \eqref{tbe} leads to the following inequality for the potential 
\begin{align}
\label{noet}
    |V'|> C V^\frac{3}{2},
\end{align}
for some constant $C$. Interestingly, the constraint $|V'|>C V^\frac{3}{2}$ is also the standard condition for no-eternal inflation \cite{rudelius2019conditions} (see \cite{Blanco-Pillado:2019tdf} for an alternate scenario which is able to provide eternal inflation even if this condition is not satisfied). This is consistent with the results of figure \ref{regions-gt}, where we can see that the TCC-allowed region excludes the eternal inflation locus represented as a black vertical line.

It is worth noting that the setup is generally sensitive to $\mathcal{O}(1)$ factors which get hidden on the value of the constant $C$.  For example, the actual curve of $\theta=1/2$ in figure \ref{regions} gets a logarithmic correction $\theta=1/2+\frac{\log(C)}{\log\Lambda}$ if we keep track of $C$ in calculating $\theta$, where $C$ can even get a mild dependence on $\Lambda$. Hence, eternal inflation is only marginally ruled out, and some models with a large enough constant $C$ might still be allowed. We will discuss this in more detail in subsection \ref{sec:o1}.

Reference \cite{rudelius2019conditions} also proposed that eternal inflation might be in the Swampland; here, we have derived this condition on the effective potential from applying TCC to metastable de Sitter vacua. As explained in \cite{rudelius2019conditions},  a metastable dS scenario is only compatible with eternal inflation if $\Gamma/H^4\leq \mathcal{O}(1)$; this is the exact opposite of what TCC requires. We have also seen that this condition maps exactly to the usual $V'\gtrsim V^{3/2}$ for the effective potential. This is evidence that the dual description we have constructed correctly captures the physics and it is a non-trivial consistency check for our computations. 

 This relation between TCC and no-eternal inflation is actually intriguing from the perspective of the effective potential, since as shown in subsection \ref{MVQ} there is no obvious a priori reason why the
TCC should imply \eqref{noet}. This result comes about only when we include membranes in the picture.
One might have tried to show that TCC forbids eternal inflation by arguing that if inflation is eternal, there will be some patch where Planckian modes will be stretched beyond the Hubble horizon, naively leading to a violation of TCC, and thus, to the conclusion that TCC forbids eternal inflation. There are two problems with this naive argument:\begin{itemize} \item To violate TCC, an inflationary patch with a homogenous Hubble parameter must contain a mode as it goes from Planckian to Hubble size. Such a patch does not typically exist in eternal inflation since bubbles of true vacuum are constantly appearing. \item Since inflation lasts forever, one could argue that all sorts of unlikely things will happen somewhere eventually, including a TCC-violating Hubble patch. This illustrates that the current formulation of TCC is a semiclassical statement in terms of expectation values of quantum operators that only deals with what happens ``on average'', and it might be violated statistically, like the second law of thermodynamics, and point towards a  more fundamental quantum mechanical version of TCC that is absolute. 
\end{itemize}

To sum up,  TCC is a statement about the overall shape of the potential, but assuming the potential effectively describes tunneling between nearby vacua, we can get an additional point-wise result which implies eternal inflation is marginally ruled out. 

\subsection{Subleading corrections}\label{sec:o1}
Throughout most of this paper, we have been cavalier regarding $\mathcal{O}(1)$ factors and other subleading corrections. For instance, we have neglected the $\log(1/H)$ logarithmic term in the TCC bound, or the WGC extremality factor in \eqref{WGC0}. The reason for this is that we cannot compute some of these in complete generality, such as $\mathcal{O}(1)$ corrections to the prefactor in \eqref{P}. Although the qualitative results and conclusions we present in this paper are insensitive to such subleading corrections, they become important when determining the fate of effective potentials satisfying 
\beq
|V'|\sim C\, V^{3/2}\ .
\label{et2}
\eeq
The region near the tip of the TCC-allowed blue triangle in figure \ref{regions-gt} is sensitive to these numerical factors, and might get extended to cross the vertical line at $\theta=1/2$ marginally allowing potentials satisfying \eqref{et2} for a certain value\footnote{The proposed values for C  in the condition for eternal inflation in the literature varies, e.g. $C=\frac{1}{\sqrt 2 \pi}$ in \cite{rudelius2019conditions} and $\frac{1}{2\pi\sqrt{3}}$ in \cite{page1997space}.} of $C$.

\begin{figure}[!htb]
\begin{center}
\includegraphics[width=0.48\textwidth]{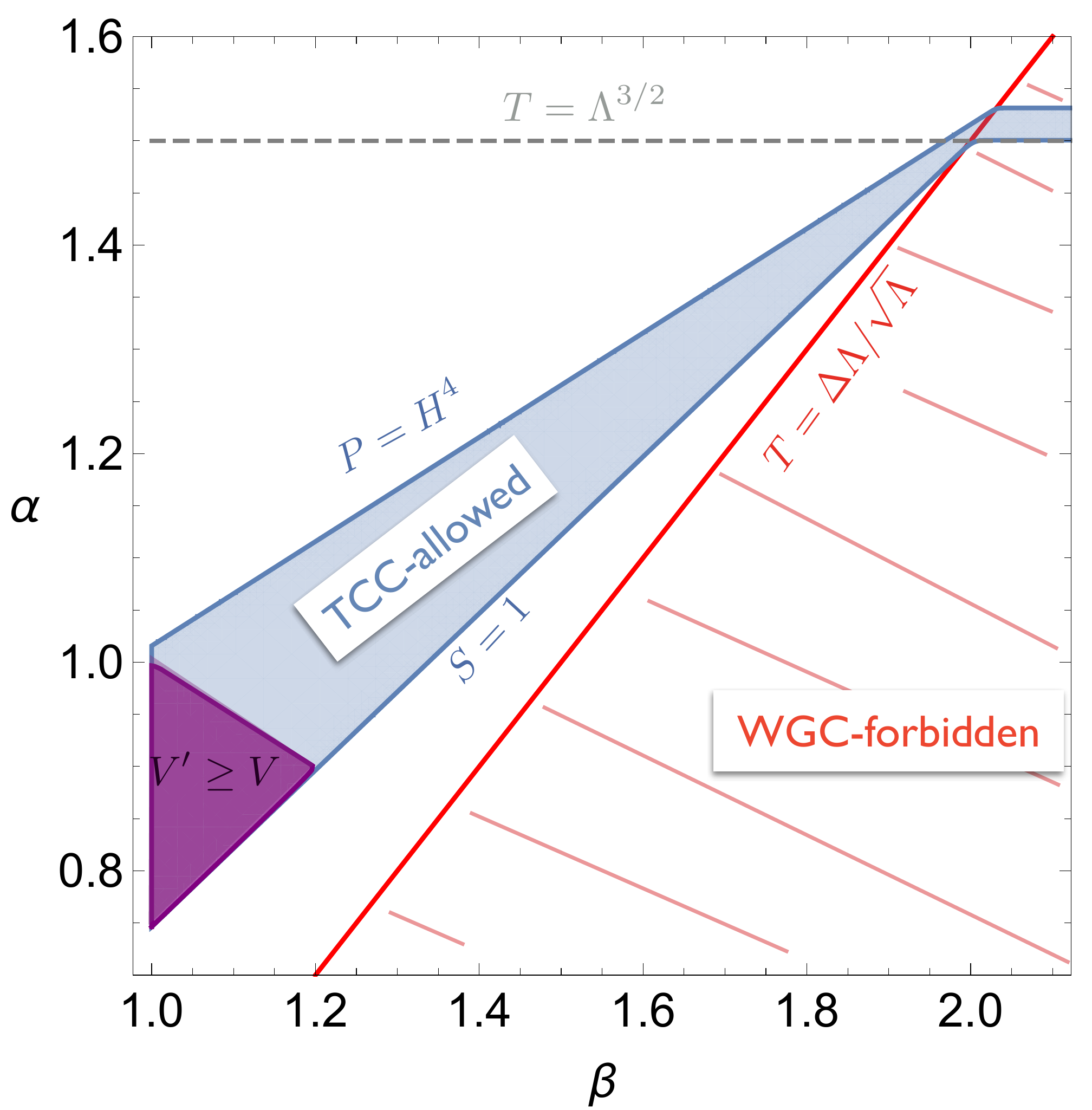}
\includegraphics[width=0.48\textwidth]{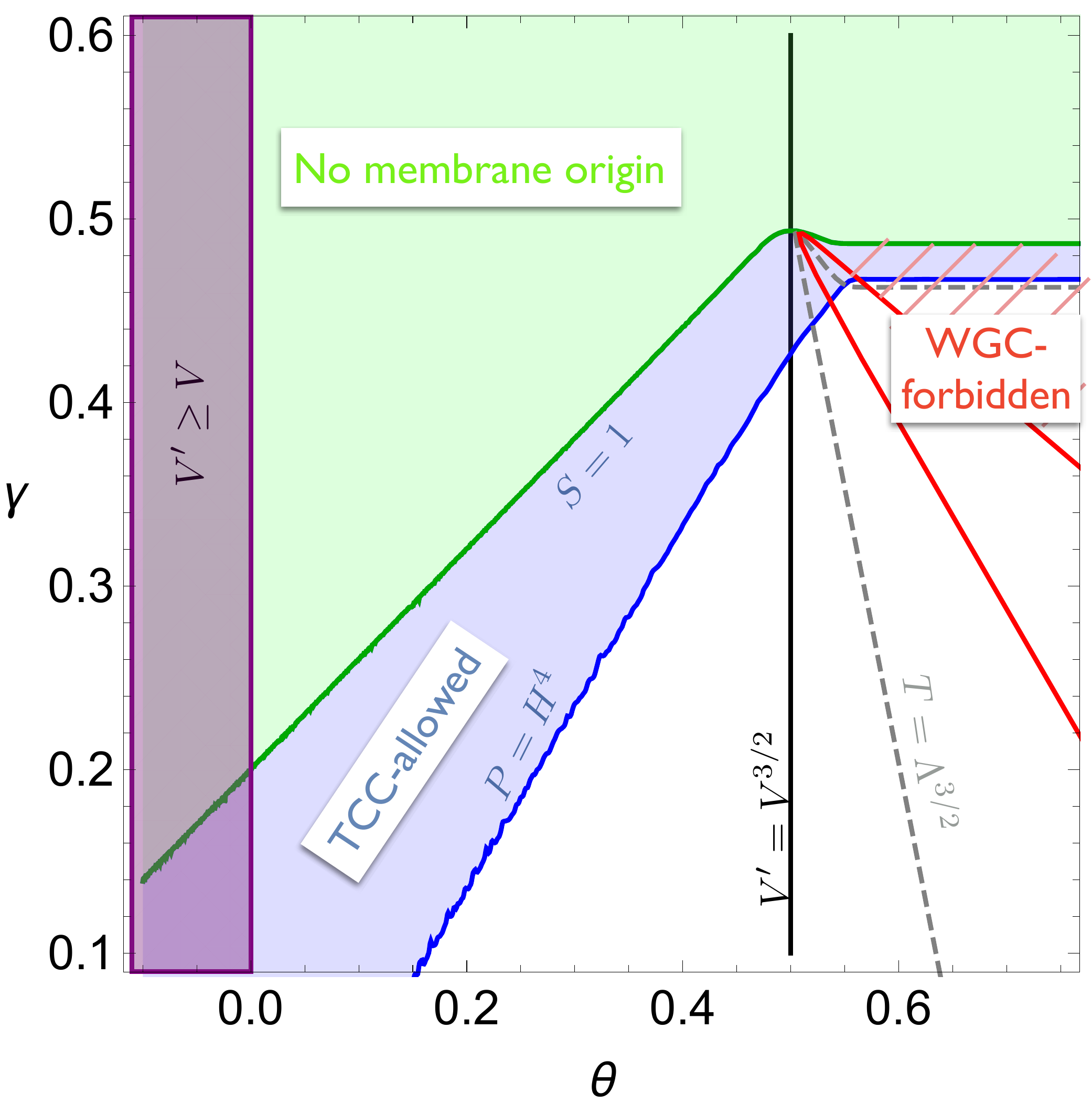}
\end{center}
\vspace{-0.7cm}
\caption{Allowed regions in the $(\alpha,\beta)$ plane (left panel) and $(\gamma,\theta)$ (right panel) for $\Lambda=10^{-120}$  taking into account the logarithmic correction on the TCC.  The light blue TCC region now grows an extra ``tube'' that makes it consistent with any value of $\theta$. The red curve correspond to WGC,  purple region corresponds to $V'/V>1$.  Eternal inflation is marginally ruled out by TCC and WGC, but not by either on its own.} \label{regions-O1}
\end{figure}

For concreteness, dS gravitational corrections to the prefactor and instanton action push the TCC-allowed region to the left, moving it away from the eternal inflation locus by introducing a negative correction to $\beta_{max}$ in \eqref{betamax}  of order $\mathcal{O}(1/|\log \Lambda|)$ that increases the value of $C$ in \eqref{noet}. Contrarily, the logarithmic term in the TCC bound, $\Delta t<H^{-1}/\log H$ implies a positive correction to $\beta_{max}$ of order $\mathcal{O}(\log(\log\Lambda)/|\log \Lambda|)$ that pushes the TCC+WGC-allowed region to the right. Depending on the exact value of this correction, the TCC-allowed region might get extended to parametrically large values of $\beta$, as illustrated in figure \ref{regions-O1}. However, the WGC will always cut this region providing, even in this case, a maximum value of $\theta$. Hence, in this case, the bound \eqref{noet} is still valid but the value of $C$ will be smaller than one and depend logarithmically on $\Lambda$. Other corrections coming from the prefactor or the WGC could also in principle push the bound in one direction or another.

In any case, we can conclude that potentials satisfying $|V'|\leq CV^{3/2}$ are forbidden by the swampland constraints for a certain factor $C$ that is sensitive to all these corrections and could have a logarithmic $\Lambda$ dependence. Therefore, any attempt to  rule out a concrete model of eternal inflation would require a better knowledge of all possible subleading corrections. This is certainly an interesting avenue to further study in the future. At the moment, we can only conclude that eternal inflation is \emph{marginally} forbidden by the swampland constraints.

\subsection{Higher dimensions}\label{sec:hig}

We showed that the TCC marginally implies WGC and the no-eternal inflation condition in 4 dimensions. One can generalize all the calculations to show the same holds in higher dimensions as well. Following is a naive computation to demonstrate how this plays out int higher dimensions. In $d$-dimensions, the equations \eqref{tcc1} and \eqref{tcc2} change to
\begin{align}
    &S\sim\frac{T^d}{(\Delta\Lambda)^{d-1}}\lesssim 1\rightarrow \alpha\geq\frac{d-1}{d}\beta,\nonumber\\
    &P\sim\frac{T^d}{(\Delta\Lambda)^{d-2}}\gtrsim H^d\rightarrow\alpha\leq \frac{d-2}{d}\beta+\frac{1}{2}.
\end{align}
These two lines constraints together imply $\alpha\geq\beta-\frac{1}{2}$ which is the WGC. Moreover, the above inequalities set an upper bound $\frac{d}{2}$ on $\beta$. Plugging that upper bound into \eqref{V} leads to 
\begin{align}
   \left (\frac{|V'|}{V}\right)^2\sim\Lambda^{\beta-\frac{3}{2}}\Gamma \mathcal{V}_{\rm eff}\gtrsim \Lambda^{\beta-1} \gtrsim\Lambda^{\frac{d}{2}-1},
\end{align}
where we used TCC in the second equation. We can write the above inequality as $|V'|\gtrsim V^\frac{d+2}{4}$ which is the no-eternal inflation condition in $d$ dimensions\footnote{The no eternal inflation condition derived in \cite{rudelius2019conditions} could be generalized to higher dimensions as follows. In higher dimensions, the Fokker Planck equation (2.11) in \cite{rudelius2019conditions} takes the form $\dot P[\phi,t] = A\partial_i\partial^i P[\phi,t]+B\partial_i((\partial^i V(\phi))P[\phi,t])$ where $A\sim H^{d-1}$ and $B\sim H^{-1}$. This modifies the Gaussian solution (3.7) in \cite{rudelius2019conditions} to $Pr[\phi>\phi_c,t]\sim\exp[-\frac{t}{\sigma^2}]$ where $\sigma\sim H^\frac{d+1}{2}/|V'|$. In order to have eternal inflation, the Hubble expansion must beat this exponential decay i.e. $H\gtrsim \frac{|V'|^2}{H^{d+1}}$. This results in the no eternal inflation condition $|V'|>KV^\frac{d+2}{4}$ for some constant $K$ which depends on $\mathcal{O}(1)$ factors in computation of $A$ and $B$. }. Therefore, we find TCC marginally implies WGC and no-eternal inflation condition in all higher dimensions as well. This points to a deeper relationship between TCC and WGC as this result holds in all dimensions and not just 4.

\section{Cosmological implications}

In this section, we study the cosmological implications of our results assuming that the relevant potentials are dual to a fast decaying dS. In particular, we are interested in the consequences of our results for the emergent inflationary models and the dark energy. 

\subsection{Inflation}

In \cite{bedroya2020trans} it was shown that the simplest TCC-compatible potential that could fit the observations such as the CMB power spectrum is an inverted parabola.  A similar hilltop model was discussed in \cite{schmitz2020trans}.  A concrete example of this can be taken to be $V=V_0(1-0.02\phi^2)$ defined over $[\phi_i,\phi_f]$ where $\phi_f$ is fixed by observation to be
\begin{align}
    \phi_f\simeq 3.9\times10^5\cdot\left(\frac{V_0}{M_{pl}}\right)^{0.505}.
\end{align}
Plugging this into the potential gives 
\begin{align}
    |V'|(\phi_f)\simeq 8\times10^3\, V_0^{1.505}.
\end{align}
For small enough $V_0$ and large enough $\phi_i$ the above potential is consistent with the no-eternal inflation condition as well as the TCC. Even though the potential is consistent with the TCC, it still suffers from a severe fine-tuning problem due to its short-field-range. This is because the field range is not long enough that a generic trajectory converges the slow-roll attractor. This initial condition problem seems to be an unavoidable consequence of the TCC for inflationary models \cite{bedroya2020trans}. As discussed in \cite{bedroya2020} there is an additional fine-tuning problem that goes back to the freedom in choosing the dS vacuum among the $\alpha$-vacua. The only $\alpha$-vacuum that can produce the scale-invariant CMB fluctuations is the Bunch-Davis (BD) vacuum. It was argued that if the dS space lives long enough the fine-tuning problem goes away because any $\alpha$-vacuum will thermalize into the BD vacuum \cite{kaloper2003initial}. This argument does not apply to TCC-compatible dS spaces due to their short lifetime. 

\subsection{Dark Energy}

Suppose the evolution of the cosmological constant is given by a scalar field whose potential comes from the successive short inter-vacua tunneling as discussed in this paper. As the scalar field rolls down, the characteristics of the domain wall corresponding to the potential evolve. We can view the rolling of the scalar field as a trajectory in the $\alpha-\beta$ plane in the membrane picture. TCC tells us that in the asymptotic of the field space $\frac{|V'|}{V}\gtrsim\mathcal{O}(1)$. Thus, the trajectory in the $\alpha-\beta$ plane (figure \ref{regions}) eventually reaches the purple curve. In fact, this must happen within a TCC time. This is because before we hit the purple curve the potential is very flat ($|V'|\ll V$) and the Hubble parameter is almost constant. From observations we know that the equation of state parameter $w$ is close to $-1$ which means the quintessence potential is not steep, i.e. $|V'|\lesssim V$. This leaves two possibilities for the current state of our universe:  we are very close to the purple curve where the potential begins to steep down, or we are still wandering in the blue region with a plateau potential while moving toward the purple curve. 
 
 \textbf{Case 1: Near the $|V'|\sim V$ curve}

In that case, the universe while remaining in the blue region must be close to the purple curve. 
That means we are close to the line that connects 
$(\alpha,\beta)\simeq(1,1)$ to $ (\alpha,\beta)\simeq (0.9,1.2)$. All these points correspond to the same slope $\frac{|V'|}{V}\sim\mathcal{O}(1)$, however they differ in the scale of the bounce radius. From the equation \eqref{S1} we find $R\simeq R_0\sim \Lambda ^{\alpha-\beta}$.
This gives a range $1\lesssim R\lesssim \Lambda^{-0.3}$ for the bounce radius in Planck units. After restoring the Planck length, it implies $l_{pl}\lesssim R \lesssim 10^{35} l_{pl}\sim 1\, \text{m}$.

This scenario is also phenomenologically appealing as it provides a cosmological relaxation mechanism to generate a small cosmological constant, consistent with the current expansion of our universe. As long as we are close the to the purple boundary with $V'/V\sim \mathcal{O}(1)$, it is just a matter of time to lower the cosmological constant by bubble nucleation to a very small value, even if the initial value at the beginning of the cosmological evolution was very big. This is very similar to the dynamical neutralization of the cosmological constant by Brown-Teitelboim and Bousso-Polchinski, whenever we have a landscape of flux vacua. If we are close to $\beta=1$, the variation of vacuum energy $\Delta\Lambda$ becomes very close to $\Lambda$, so after a few transitions, one would end up with a very small value for the cosmological constant. The drawback is that the effective description in which we can average over the discrete jumps breaks down and one would need a very finely grained landscape in order not to miss our current tiny value of $\Lambda$. In any case, regardless of where we are in the $(\alpha,\beta)$-plane, as long as it is close to the purple region, it is always possible to find some effective potential that reaches a tiny value of the cosmological constant in less than the Hubble time since the whole blue region is consistent with the TCC. This scenario could also explain the cosmological coincidence problem if something drastic happens when we reach a small value of $\Lambda$. A tantalizing possibility is that the effective field theory drastically breaks down precisely when getting a small value of $\Lambda$ and entering into the purple region, because we could get then access to transplanckian field ranges and infinite towers of states should become light according to the SDC.

 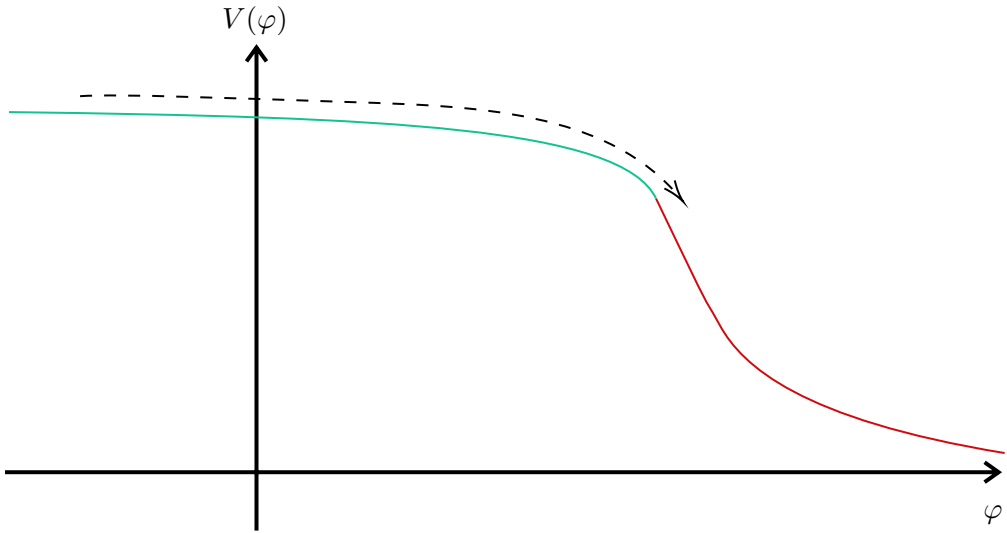
\begin{figure}
     \centering

\tikzset{every picture/.style={line width=0.75pt}} 

\begin{tikzpicture}[x=0.75pt,y=0.75pt,yscale=-1,xscale=1]

\draw [line width=1.5]  (110,240.14) -- (606,240.14)(235.56,26.5) -- (235.56,269.5) (599,235.14) -- (606,240.14) -- (599,245.14) (230.56,33.5) -- (235.56,26.5) -- (240.56,33.5)  ;
\draw [color={rgb, 255:red, 208; green, 2; blue, 27 }  ,draw opacity=1 ]   (435,102.5) .. controls (466,167.5) and (456,146.5) .. (467,166.5) .. controls (478,186.5) and (506.79,213.57) .. (609,230.5) ;
\draw [color={rgb, 255:red, 1; green, 118; blue, 255 }  ,draw opacity=1 ]   (112,59) .. controls (298.95,61.4) and (418,66.5) .. (435,102.5) ;
\draw  [dash pattern={on 4.5pt off 4.5pt}]  (147.5,51) .. controls (166.83,49.7) and (222.5,52) .. (283.5,54) .. controls (343.89,55.98) and (411.14,57.96) .. (447.41,102.63) ;
\draw [shift={(448.5,104)}, rotate = 231.95] [color={rgb, 255:red, 0; green, 0; blue, 0 }  ][line width=0.75]    (10.93,-3.29) .. controls (6.95,-1.4) and (3.31,-0.3) .. (0,0) .. controls (3.31,0.3) and (6.95,1.4) .. (10.93,3.29)   ;

\draw (217,4.4) node [anchor=north west][inner sep=0.75pt]    {$V( \varphi )$};
\draw (597,255.4) node [anchor=north west][inner sep=0.75pt]    {$\varphi $};

\end{tikzpicture}

     \caption{The graph shows a generic effective potential that would emerge from successive inter-vacua tunnelings. In the plateau part of the graph which is drawn in green, we have $|V'|\ll V$. This corresponds to part of the cosmic evolution spent in the blue region of the figure \ref{regions}. The steep part of the potential drawn in red has $|V'|\gtrsim V$. This corresponds to the part of the cosmic evolution spent inside or near the purple region in the figure \ref{regions}. }
     \label{generic}
 \end{figure}
 
\textbf{Case 2: Far from the $|V'|\sim V$ curve}

Suppose our universe in the blue region the $\alpha-\beta$ plane in figure \ref{regions} sufficiently far from the purple curve. In that case, we have $|V'|\ll V$ which means the potential is so flat that effectively behaves like a cosmological constant. This is consistent with observations. Moreover, the TCC is also satisfied because as discussed in subsection \ref{MVQ}, the only constraint that TCC imposes on nearly flat potentials is that the age of the universe must be less than $\frac{1}{H}\ln(\frac{1}{H})$ which is true in our universe. The drawback is the usual naturalness problem of the cosmological constant since one would need to start originally with a very tiny cosmological constant as the membrane nucleation transitions will not modify its value in a significant way.

\section{Conclusions}
In this paper, we applied some of the Swampland conjectures to short-lived de Sitter spaces and described the resulting decay in terms of an effective theory.
 We saw that WGC and the generalized distance conjecture lead to a restricted region in parameter space which surprisingly includes the full region implied by the TCC.  In other words, TCC in this context seems to ``know'' about WGC and the generalized Distance Conjecture.  This relationship between different Swampland conjectures which is frequently encountered, reinforces the belief in their validity.
In studying the resulting effective theory, we found that they lead to a scalar field with a restricted type of potential.  In particular, even though a potential allowing eternal inflation is consistent with TCC, the resulting potentials we obtain from dS membrane picture are marginally inconsistent with eternal inflation.  However eternal inflation is not completely ruled out by our considerations and there is a small region in parameter space that may in principle allow it depending on subleading corrections we have neglected in our analysis. This is an interesting question that should be explored in the future.

Our results apply to situations that can be described as a cascade of non-perturbative nucleation of bubbles in de Sitter space. We have constructed an effective potential that provides a dual description of the low-energy physics of the cascading membranes, connecting the membrane and cascade pictures. Our methods can be extended to a variety of interesting situations, e.g. when the membranes are charged under multiple 3-form gauge fields or the membranes are unstable and can grow holes in their surface corresponding to strings magnetically charged under axions coupled to the 3-form gauge fields.  

So how general is the membrane picture? On a more speculative note, we could take the radical position that  \emph{any} quasi-dS potential consistent with string theory always admits such a membrane origin. If so, the conclusion that eternal inflation is in the Swampland, which we got here from the membrane picture, would be general. In particular, near the infinite distance boundaries of the field space in string theory compactifications, the scalar potential is known to exhibit a runaway behavior towards the asymptotic limit that typically satisfies the de Sitter conjecture. Such a runaway could be explained if it is actually the effective description of a cascade of membrane nucleation with a tension approaching the region $V'\geq V$ in figure \ref{regions}. To determine whether this is a sensible scenario, we would need a better understanding of possible constraints on the dynamics on the $(\alpha,\beta)$-plane in addition to the ones studied in this paper.

It would be interesting to see if other Swampland conjectures can also be brought to play in this context.  For example, the cobordism conjecture  \cite{McNamara:2019rup} predicts that there is always a bubble of nothing in a quantum theory of gravity.  What is the relation of our ``minimal bubble'' to the bubble of nothing?  Can this place an upper bound on the lifetime of dS which is even stronger?  Given the importance of a deeper understanding of dark energy for the future evolution of our universe, it is worthwhile pursuing aspects of short-lived dS from all possible perspectives.

\textbf{Acknolwedgements} We thank Prateek Agrawal for valuable discussions. We have
greatly benefited from the hospitality of UC Santa Barbara KITP where this project was initiated in the workshop ``The String Swampland and Quantum Gravity Constraints on Effective Theories.'' This research was supported in part by a grant from the Simons Foundation (602883, CV).

\appendix

\section{Subtleties of the thin-wall approximation}\label{app:CdL}
In the main text we presented a simplified version of the thin-wall discussion to make the presentation easier to follow, but in such a way that the main conclusions are unaltered. The actual calculations are more complicated, and we discuss them here, in order of appearance in the main text.
\subsection{Gravitational effects in thin-wall formulae}\label{Gammathin}
In the discussion in section \ref{sec:CdL}, we neglected gravitational effects that are relevant when the bubble radius is comparable to the Hubble scale. This turns out a posteriori to be a good approximation since the results only change by an $\mathcal{O}(1)$ factors, but one needs to check the full result, which we do here. 

Taking into account gravitational effects, the euclidean bounce solution \cite{Ibanez:2015fcv, Garriga:1993fh} is given by
\begin{align}
S=2\pi^2Tr^3+\frac{12\pi^2}{\kappa^2}\left\{\frac{1}{V_f}\left[(1-\frac13\kappa V_f r^2)^{\frac32}-1\right]-\frac{1}{V_i}\left[(1-\frac13\kappa V_i r^2)^{\frac32}-1\right]\right\}.\label{bounce}
\end{align}
where $T$ is the tension of the membrane (the wall), and the initial and final vacuum energies are $V_i$ and $V_f$ respectively, with $V_f<V_i$ (we will discuss the possibility of up-tunneling below). The critical radius corresponds to the value of $r$ that minimizes the above action, which implies solving the following equation,
 \begin{align}
 \gamma r=- T  \sqrt{1-r^2 \Lambda _i},\quad \gamma\equiv \left(\frac{ T^2}{4}+\Lambda_f-\Lambda_i\right).\label{sasas}
 \end{align}
We can see that non-trivial solutions exist only for $\gamma<0$, namely for bubble radius smaller than the de Sitter horizon $r<\Lambda_i^{-1/2}=H_i^{-1}$ where $H$ is the Hubble scale. We have defined $\Lambda\equiv \kappa V/3$ and set $\kappa=1$ to work in Planck units. Notice that the case $\gamma=0$ corresponds to a bubble of Hubble radius, while $\gamma<0$ implies the following critical radius
 \begin{align}
 R^2=\frac{1}{\left(\frac{\gamma}{ T}\right)^2+\Lambda_i}\label{rsquared}
 \end{align}
Plugging this back into \eqref{bounce} one gets the final result for the action $S$. For later convenience, it is useful to define the parameters,

 \begin{align}
 \label{pq}
 p\equiv\frac{T}{\sqrt{\Lambda}}
\ ,\quad q\equiv\sqrt{\Lambda}\left(\frac{ T}{\Delta\Lambda}\right)=R_0H\end{align}
where we have renamed $\Lambda\equiv \Lambda_i$ and $R_0\equiv T/\Delta \Lambda$ is the critical radius of a bubble in flat space. When the parameter $p$ becomes small, gravitational corrections become subleading, and we can expand the instanton action for small $p$ to obtain:
 \begin{align}
 \label{action1} 
S=w(q)\frac{ T}{  \Lambda^{3/2}}+\mathcal{O}(p^2),\quad \quad \frac{w(q)}{2 \pi ^2}=\frac{1+2/q^2}{\sqrt{1+1/q^2}}-\frac{2}{q}
\end{align}
 In this limit, the bubble radius reads
 \begin{align}
R^2\simeq \frac{1}{\Lambda}\frac{1}{1+(1/q)^2}\ \rightarrow \ (RH)^{-2}\simeq 1+(R_0H)^{-2} \label{wuhu}\end{align} 
 Hence, the size of the bubble is parametrised by the value of $q$, which yields two limits of interest.  On one hand, if $q\ll 1$, the critical radius is small compared to the de Sitter horizon and we recover the result for the transition rate in the flat space limit,
 \beq\label{S1}
S\simeq \frac{T}{\Lambda^{3/2}}q^3=\frac{T^4}{\Delta\Lambda^3}\equiv S_0 \ , \quad R\simeq R_0=\frac{T}{\Delta\Lambda}
\eeq
On the other hand, if $q\simeq \mathcal{O}(1)$, the bubble radius is of Hubble size and we get
 \beq\label{S2}
S\simeq \frac{T}{\Lambda^{3/2}}\ ,\quad R\simeq H^{-1}
\eeq
Notice that $q=1$ is the largest value that this parameter can take which is still consistent with a solution to \eqref{sasas}, so \eqref{S2} gives the largest possible radius and the smallest possible euclidean action of an instanton solution describing bubble nucleation in de Sitter space in the thin wall approximation. 

We also comment briefly on the prefactor \cite{Garriga:1993fh}. The full expression taking into account gravitational backreaction is
\begin{equation}
\label{prefactor1}
P=\frac{e^{-\zeta'_R(-2)}}{4}T^{2}R^{2}\simeq \frac{T^2R_0^2}{1+(R_0H)^2}\sim\frac{S^{2}}{R^4} 
 \end{equation}
where the last equality is true modulo a $\mathcal{O}(1)$ function of $HR$ only which goes to 1 at zero, and $\zeta'_R(-2)=-0.0394\ldots$. 

A precise determination of the prefactor outside of the thin-wall approximation requires the calculation of a one-loop determinant around the Euclidean saddle, and it is both complicated and detail-dependent (see \cite{Dunne:2005rt} for an example). As shown in \cite{Garriga:1993fh}, in the thin-wall approximation it is possible to determine the prefactor since the only low-energy degrees of freedom that contribute to the prefactor are fluctuations of its local position (fluctuations of the Goldstone associated to translational invariance). These will have energies of the order of $1/R$, while we will assume that the next excitation, corresponding to internal worldvolume degrees of freedom, will appear at a much higher energy scale. %

Even though this computation only took into account Goldstone modes, we expect our conclusions to hold modulo $\mathcal{O}(1)$ corrections if a finite number of worldvolume degrees of freedom at the scale of the Goldstones are included. For instance, if the domain wall is a D-brane, we would expect to have worldvolume gauge fields and gauginos as well. Since we are not sensitive to $\mathcal{O}(1)$ coefficients, we will drop it in \eqref{prefactor1}.

Combining \eqref{prefactor1} and \eqref{action1} we get that the transition rate per unit time and volume is given by
\beq
\label{Gamma}
\Gamma= H^4\left(\frac{T}{H^3}\right)^2\frac{(R_0H)^2}{1+(R_0H)^2}\, \exp\left({-\frac{ T}{  H^{3}}w(R_0H)}\right)
\eeq
The above thin-wall expressions can be easily generalised to any dimension \cite{Garriga:1993fh}. We use this generalization in subsection \ref{sec:hig}.

\subsection{Up-tunneling}\label{sec:up}
In de Sitter space, up-tunneling is allowed due to the gravitational effects, but it is more suppressed than down-tunneling. More precisely, as discussed in \cite{Garriga:1993fh}, up-tunneling is described by the same kind of CdL instanton that down-tunneling, considering an anti-membrane rather than a membrane. In four dimensions, the difference in action between these two tunneling rates is
 \begin{equation}S_{\text{up}}-S_{\text{down}}\propto \frac{\Delta \Lambda}{\Lambda^2},\label{avv}\end{equation}
 which will be large in most of our parameter space, but can be significant near the eternal inflation point. Notice that the difference \eqref{avv} is also essentially the difference between the entropies of the down-tunneling and up-tunneling de Sitters. The reason up-tunneling is suppressed is entropic. 
 
Uptunnelling is responsible for the last term in the formula \eqref{fulexp} for the effective potential. It is never dominant in the region allowed by the swampland conjectures.

\subsection{Regime of validity of CdL formulae}
In most of the region of interest to us, the action of the Euclidean instanton \eqref{flat} is small. However,  the usual lore is that semiclassical expressions such as these are only accurate as long as the instanton action is large and so there is exponential suppression. So how come that we can use it more generally?  CdL is essentially an application of the WKB formula to field theory, and this is controlled not by whether there is exponential suppression or not, but by whether the perturbation parameter is small. These two notions can differ in a theory that has more than one parameter.

An illustrative example is Schwinger's original calculation of the decay via emission of charged pairs in (1+1) dimensions. Schwinger obtained a vacuum decay amplitude (later reinterpreted as a pair production rate \cite{Nikishov:1970br,Cohen:2008wz}) given by
\begin{equation} \Gamma=\frac{(qE)^2}{4\pi^3}e^{-\frac{\pi m^2}{qE}}.\label{schw}\end{equation}
This calculation was done in the semiclassical limit $q\rightarrow0, E\rightarrow \infty$, with $qE$ fixed. The small parameter is therefore the electron charge $q$ and we can expect that \eqref{flat} is just the first in a series of corrections suppressed in higher powers of $q$. The classical instanton action is becoming small when $m\rightarrow 0$, yet the result is still trustworthy because the expansion parameter is $q$, which remains small. 

This is not always the case. When computing the potential generated for the $\theta$ parameter in a Yang-Mills theory, there is only one small coupling, the Yang-Mills coupling $g$. The instanton action $S=8\pi^2/g^2$ only depends on $g$, and sending $g\rightarrow\infty$ means both that the instanton action becomes small and the perturbative expansion fails.  

The CdL scenario is very similar to the Schwinger example above. Instead of the particle mass, we have the tension $T$, and the parameter $qE$ in the Schwinger model, which is just the difference in vacuum energy before/after pair nucleation, is replaced by $\Delta \Lambda$ in our example. The change in vacuum energy $\Delta \Lambda$ is then related to the  background flux density in the false vacuum as
\begin{equation}\Delta \Lambda= g_3^2\, n,\end{equation}
where $n$ is an integer parametrizing the background ``top-form flux density''. The parameter $g_3$ is the 3-form gauge coupling \cite{Bousso:2000xa} which controls the strength of interactions and backscattering between the domain walls.

The thin-wall computations above should be understood as taking place in a formal limit where $g_3$ is going to zero and $n$ diverges in such a way that
\begin{equation}\Delta \Lambda= g_3^2 n\rightarrow\text{const}.\end{equation} 
More physically, this should be thought of as a limit in which brane-brane interactions are switched off, but branes still respond to the background difference in vacuum energies. In such a limit, just as in the Schwinger case, we expect to be able to trust the thin-wall expression even when the membrane tension $T$ is small and there is no exponential suppression since it is just the leading piece of the small $\Delta \Lambda$ expansion. In this case, the physics is dictated by the prefactor. But we also emphasize that this is an assumption we make and which we cannot prove. Ultimately, the reason for this is that the rigorous argument for the Schwinger effect above relied on the fact that we have a Lagrangian description of the system, which we are lacking in the higher-dimensional case since extended objects are an essential ingredient\footnote{We could try to compactify to (1+1) dimensions to recover a Lagrangian description, or we could resort to the effective action for a probe brane/particle, which is available also in higher dimensions. But none of these arguments are conclusive since the probe brane approach we do not know how to compute corrections systematically, and compactification to (1+1) would involve talking about wrapped branes, with very different kinematic properties.}.

\section{Derivation of the effective potential}\label{app:potential}
Here we discuss in detail the derivation of the effective potential introduced in the main text. 
The basic quantity we are interested in is
\begin{equation}dN/dt,\end{equation}
 the number of membranes that hit an observer per unit time. We will now do so by geometric means, as in \cite{Kashyap:2015lva}. 

\begin{figure}[!htb]\begin{center}
\includegraphics[width=0.3\textwidth]{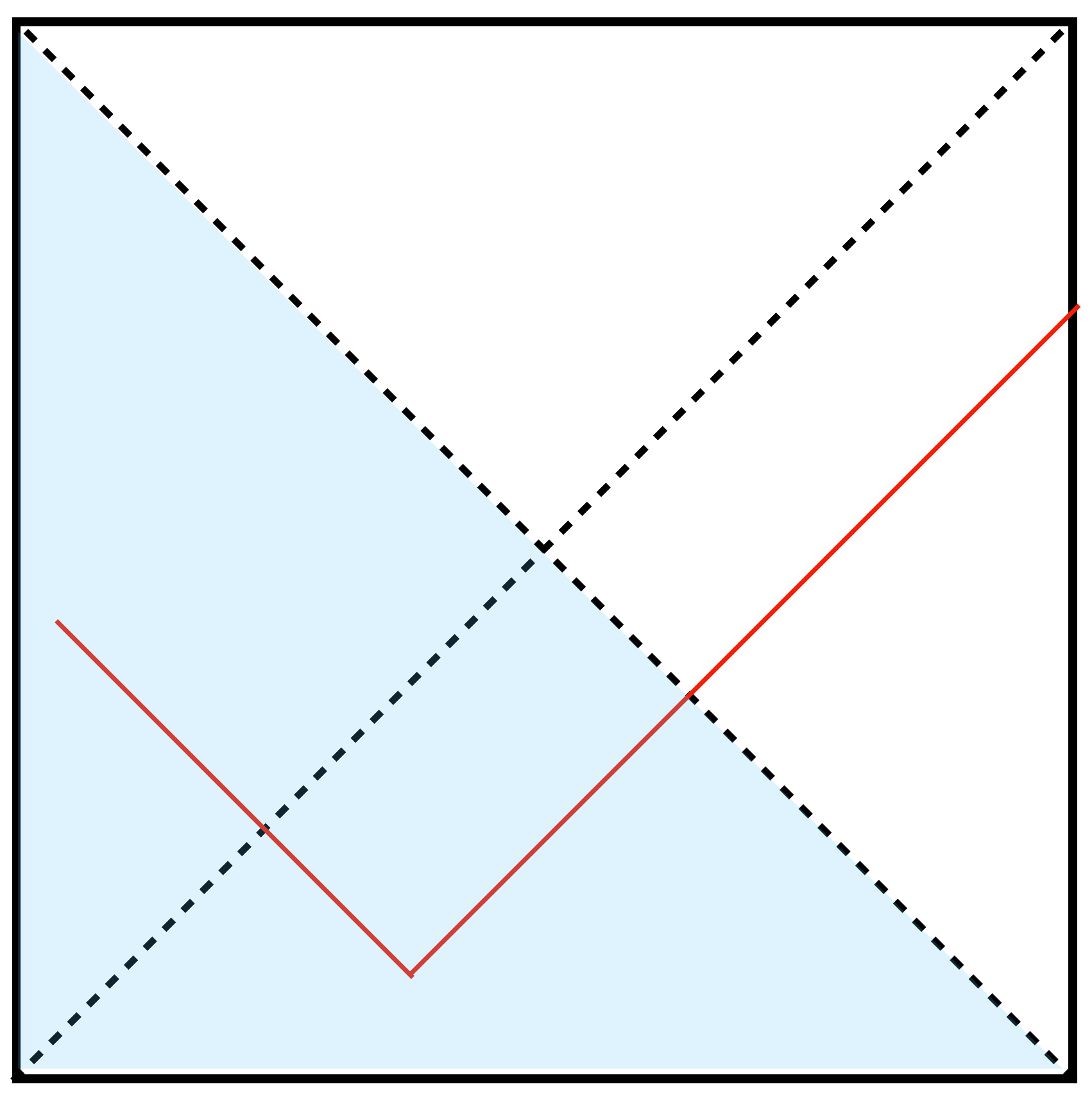}
\caption{A depiction of the conformal diagram of de Sitter space. The flat coordinates \eqref{open} only cover half of the spacetime (they do not cover the part shaded in blue). Global coordinates \eqref{dsglobal} cover all of spacetime. Surfaces of constant $\chi$ would be horizontal lines in the picture, while surfaces of constant $\tau$ asymptote to the past cosmological horizon of the patch they cover (the top left to down right diagonal line).}
\label{f0}\end{center}
\end{figure}

Take $d$-dimensional de Sitter space in conformal flat slicing:
\begin{equation}ds^2=\frac{1}{\tau^2}\left(-d\tau^2+\sum_{i=1}^{d-1} dx_i^2\right),\label{open}\end{equation}
where the coordinates $x_i$ parametrize flat space. Here $-\infty<\tau<0$ parametrizes half of a $dS_d$ (see Figure \ref{f0}).  If the observer sets her clock such that $t=0$ is at $\tau=-1$, then in general her proper time is related to $\tau$ as
\begin{equation} \tau=-e^{-t}.\end{equation}
 Let
\begin{equation} \xi =\frac{dN}{dt}=-\tau \frac{dN}{d\tau}\end{equation}
be the number of bubbles that hit the worldline of a timelike static observer at $\vec{x}=\text{const}$. Our task is to determine $\xi$. We will do this by calculating $\xi$ in two different ways, and then demanding they are both equal:\begin{itemize}
\item $\xi$ is related to the mean free path traversed by a bubble. Suppose one has a bubble whose wall is expanding at the speed of light. In the above coordinate system, it moves along a straight line $x=\tau$. When traversing a time interval $\Delta \tau$, there is a probability 
\begin{equation} P_{\text{hit}}= -\frac{\xi}{2\tau} \Delta\tau\end{equation}
that the bubble gets hit by another bubble entering its lightcone (see Figure \ref{f1}).

\begin{figure}[!htb]
\begin{center}
\includegraphics[width=0.5\textwidth]{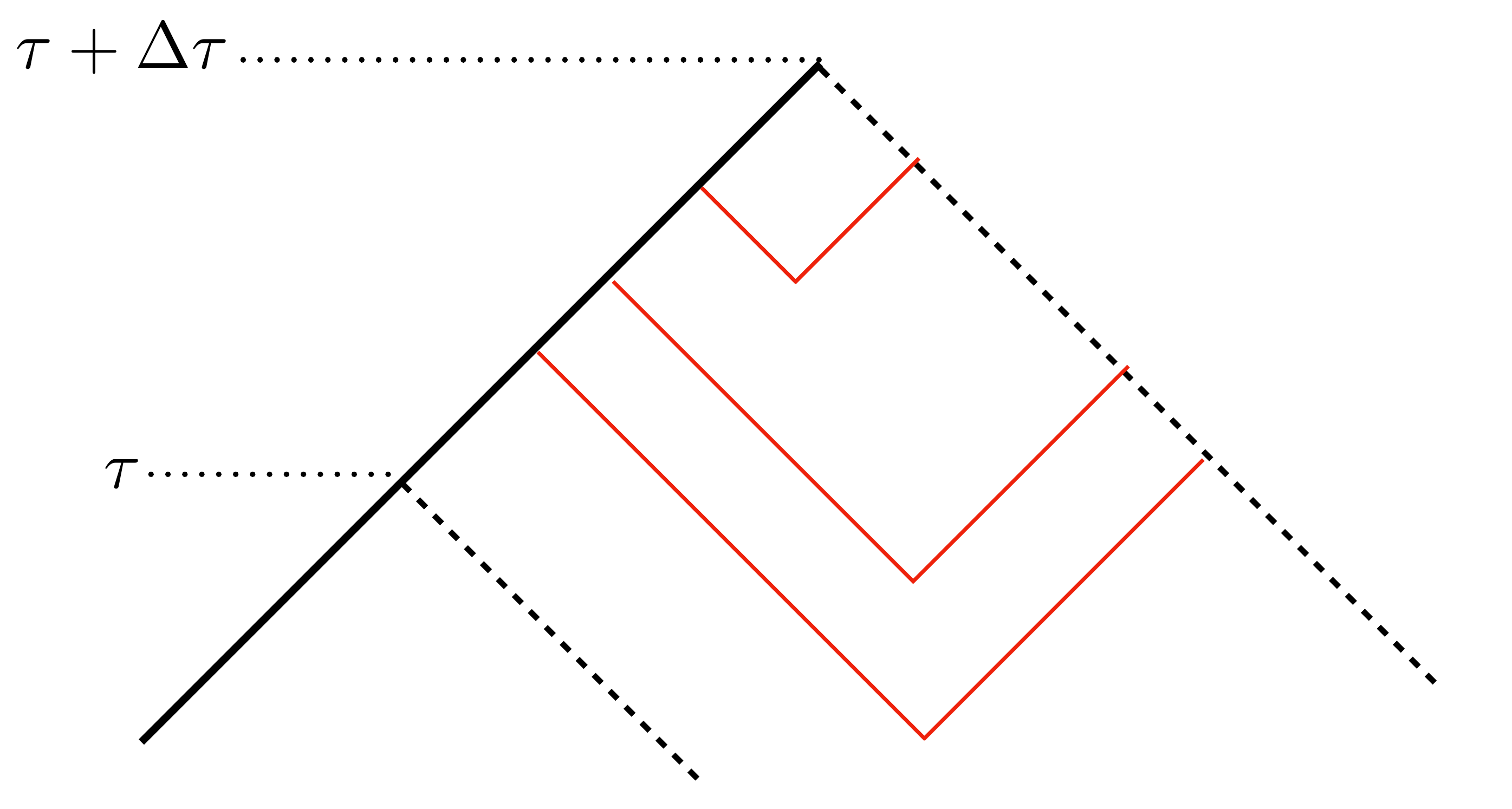}
\caption{In black, we have the worldline of an expanding domain wall. As it moves in a lightlike fashion through an interval $\Delta \tau$, there is a probability that it gets hit by a bubble. This probability depends on $\xi$, the number of membranes that arrive per unit time the worldline of static observers.}
\label{f1}\end{center}
\end{figure}

 If it gets hit, the bubble will  annihilate and disappear. The probability that the bubble survives these collisions at $\tau+\Delta \tau$, $P_{\tau+\Delta \tau}$, is equal to the probability $P_\tau$ that it made it to $\tau$, times the probability that it does not get hit by the bubble, so
\begin{equation} P_{\tau+\Delta \tau}=P_\tau+\Delta P_\tau= P_\tau \left(1-\frac{\xi}{2\tau} \Delta\tau\right),\end{equation}
which has solution
\begin{equation} P_\tau= \left(\frac{\tau_0}{\tau}\right)^{\frac{\xi}{2}}.\label{p11}\end{equation}
This equation gives the probability that a bubble which was born at conformal time $\tau_0$ actually makes it to  conformal time $\tau$ (see  \cite{Kashyap:2015lva}).  In terms of proper time, we have
\begin{equation} P_t=P_{t_0}e^{-\frac{\xi}{2}(t-t_0)},\end{equation}
which implies that the quantity $\xi$ we are actually looking for is just the inverse of the mean free path of a bubble.
\item Similarly, to compute the number of bubbles that hit an observer at conformal time $\tau_0$ in a conformal time window $\Delta \tau$, we just need to integrate the bubble production rate on the past light cone, and correct by the factor \eqref{p11} which establishes that only a fraction of the bubbles produced at conformal time $\tau$ actually make it to $\tau_0$.

Taking these two things into account, the number of bubbles we get is
\begin{equation} \Delta N = \Gamma S_{d-2} \int_{-\infty}^{\tau_0} d\tau\,\int_{r(\tau,\tau_0)}^{r(\tau,\tau_0+\Delta \tau_0)}dV\, \left(\frac{\tau_0}{\tau}\right)^{\frac{\xi}{2}},\end{equation}
where $S_{d-2}$ is the volume of $S^{d-2}$, and 
\begin{equation}r(\tau,\tau_0)=\tau_0-\tau\end{equation}
parametrizes the null radial geodesic from $(\tau,r)$ to $(\tau_0,0)$. Plugging everything back in, one gets
\begin{equation}\frac{\Delta N}{\Delta \tau_0}= \Gamma S_{d-2} \int_{-\infty}^{\tau_0} \frac{(\tau_0-\tau)^{d-2}}{\tau^d}\left(\frac{\tau_0}{\tau}\right)^{\frac{\xi}{2}}.\end{equation}
\end{itemize}
Evaluating the integral for $d=4$, one obtains
\begin{equation} -\frac{\xi}{\tau_0}=\frac{dN}{d\tau_0}=-\Gamma S_{d-2} \frac{16}{\tau_0 (\xi +2) (\xi +4) (\xi +6)},\end{equation}
which allows one to get (restoring the Hubble constant)
\begin{equation} \frac{\xi}{H}= -3+\sqrt{4 \sqrt{\frac{\Gamma S_{d-2}}{H^d} +1}+5}.\label{mfp}\end{equation}
On the other hand, we could write $\xi$ as
\begin{equation}
\label{vol}
\xi=\frac{dN}{dt}=\Gamma \mathcal{V}_{\rm eff}
\end{equation}
where $\mathcal{V}_{\rm eff}$ is some effective volume. Notice that this effective volume 
\begin{equation} 
\mathcal{V}_{\text{eff}}= \frac{\xi}{\Gamma}=\frac{H}{\Gamma}\left(-3+\sqrt{4 \sqrt{\frac{\Gamma S_{d-2}}{H^d} +1}+5}\right)
\label{Veff}
\end{equation}
is always below 1 in Hubble units. 
This works in general, but due to TCC we are interested in the regime $\Gamma/H^d\gg1$, in which \eqref{mfp} is just 
\begin{equation}\frac{\xi}{H}\sim \frac{\Gamma^{1/4}}{H^{d/4}}.\label{mfp2}\end{equation}
and the effective volume  becomes of order $\mathcal{V}_{\text{eff}}\sim \Gamma^{-3/4}$.
Equation \eqref{mfp2} can also be easily understood: in the regime where $\Gamma$ is large and bubbles are efficiently produced, a bubble will die of a collision way before it notices the expansion of the universe. As a result, it will hit another bubble by the time its volume is such that the probability of bubble nucleation is of order one, in other words $\Gamma \xi^4\sim 1$, which is precisely \eqref{mfp2}. 

Once we have the number of bubbles that hit the observer per unit time, it is a simple task to relate this to the potential. Assuming slow-roll inflation, we get
\begin{equation}\frac{dV}{dt}=\Delta\Lambda \frac{dN}{dt}=\Delta\Lambda \xi\approx \frac{(V')^2}{\sqrt{V}},
\label{dVt}
\end{equation}
We should also take into account that in dS there can be up-tunneling on top of down-tunneling, as discussed in appendix \ref{sec:up}. The effect will be small away from the curve $\beta=2$, but significant for $\beta >2$. up-tunneling membranes may annihilate with up-tunneling membranes and vice-versa, but down-tunneling and up-tunneling membranes just go through each other. As a result, to compute the change in vacuum energy, one simply has to replace \eqref{dVt} by

\begin{equation} \frac{dV}{dt}= \Delta \Lambda \frac{dN}{dt}\left(1-e^{-\Delta \Lambda/\Lambda^2}\right)\approx \frac{(V')^2}{\sqrt{V}},
\end{equation}
where the last equation is again due to slow-roll. The general effective scalar potential then reads
\begin{equation}
 \frac{V'}{V}\simeq \Delta \Lambda^{1/2} \Lambda^{-3/4}\Gamma^{1/8} \sqrt{1-e^{-\Delta \Lambda/\Lambda^2}}.\label{fulexp}
\end{equation}
This has \eqref{pot11} and \eqref{VV} as limits when the decay rate is very small or very large compared to Hubble. We use the full expression \eqref{fulexp} to make the figures in the main text.

As a final comment, we should explain why we used the open slicing \eqref{open} which, as illustrated in Figure \ref{f0}, only covers half of de Sitter space. But clearly, bubbles that nucleate in the lower half of the diagram can reach the upper half! So why don't we take them into account? The answer is related to the phenomenon of ``persistence of memory'', beautifully discussed in \cite{Garriga:2006hw,Frob:2014zka}. 
It is often stated that, due to the exponential expansion, a de Sitter universe soon ``forgets'' any initial condition; indeed, this property is crucial for the success of the inflationary mechanism. It is, however, not true in general. Certain fields, such an electric current in two dimensions \cite{Frob:2014zka} and a 3-form current in four (the case we consider here) can develop vevs that partially break the de Sitter isometries and are never diluted by the exponential expansion (any nonzero vev will do this).  While the top-form field-strength preserves all of the de Sitter isometries, the current it generates via quantum effects does not; it always picks a preferred reference frame, where e.g. it is purely spatial. This current vev in a sense retains information about what happened in the early universe, which didn't dilute away completely; hence, ``persistence of memory''.

Let us describe this in more detail. To do this, it is convenient to use conformal global coordinates, which cover all of de Sitter. In these, the metric is given by
\begin{equation} ds^2=\frac{1}{\cos^2(\chi)}\left(-d\chi^2+ d\Omega_{d-1}^2\right),\label{dsglobal}\end{equation}
where the  conformal time coordinate $\chi$ lives in $(-\pi/2,\pi/2)$ and the spatial slices are $(d-1)$ spheres. 

 Suppose the universe nucleates into existence (has a Big Bang) at some particular initial timeslice. We will take this to happen at a constant global timeslice $\chi_{\text{BB}}$, but the main lesson is actually independent of the choice of timeslice. This choice breaks the $dS$ isometries, and when computing the bubble nucleation rate, one should stop integrating at the Big Bang. The number of bubbles that reach a static observer at a later time $\chi$ (ignoring backreaction, for simplicity) can be computed as the volume of the lightcone times the nucleation rate $\Gamma$, so (in $d=4$),
\begin{equation}\frac{dN}{dt}=4\pi \Gamma \cos(\chi) \int_{\chi_{\text{BB}}}^\chi \frac{\sin^2(\chi-\chi_{\text{BB}} )}{\cos^4(\chi)}=\frac{4\pi}{3}\Gamma \frac{\sin^3(\chi-\chi_{\text{BB})}}{\cos^3(\chi_{\text{BB}})},\end{equation}
where we have stated the result again in terms of proper time $t$. Notice that, with finite $\chi_{\text{BB}}$, the result at late times is actually independent of $\chi_{\text{BB}}$ and finite,
\begin{equation} \lim_{\chi\rightarrow\pi/2} \frac{dN}{dt}=\frac{4\pi}{3}\Gamma,\end{equation}
while if we take $\chi_{\text{BB}}\rightarrow -\pi/2$ first (which would correspond to a truly eternal de Sitter, with no Big Bang), the result diverges due to the volume factor in the denominator.  This noncommutativity of limits is the technical manifestation of persistence of memory. Sending $\chi_{\text{BB}}\rightarrow -\pi/2$ first corresponds to integrating the bubble nucleation rate over all of the past light cone of a point. This is a manifestly de Sitter-invariant way of computing $dN/dt$, so of course, it must produce a dS-invariant answer; but the only dS-invariant ``values'' of $dN/dt$ are zero or infinity, so that's why the result diverges.  

By contrast, sending $\chi\rightarrow\infty$ yields a finite value; at very late times, all the information about the Big Bang has been diluted away, except for the fact that nonzero $dN/dt$ tells us that there \emph{was} a Big Bang in the first place (or, at the very least, that the system does not enjoy de Sitter invariance).  We assume this is the situation in this paper; because of the Big Bang, or for whatever other reason, there is a nonzero but finite $dN/dt$, which breaks de Sitter invariance, but the magnitude of $dN/dt$ forgets the details about the initial timeslice. This justifies using open slicing \eqref{open} and only integrating over half of the dS; it corresponds to having a Big Bang at $\tau=-\infty$, and it leads to particularly simple computations. As we have just explained, the result at late times is independent of this choice.

\section{Constraints on Hawking-Moss}\label{app:HM}

In the Hawking-Moss (HM) transition \cite{hawking1982supercooled} the universe tunnels from a local minimum to a local maximum and it happens everywhere at once. This spatial homogeneous transition can be interpreted \cite{Weinberg:2006pc} as a thermal fluctuation of a horizon-sized region up to the top of the barrier, followed by the rolling of the field down to the true vacuum. It also has an entropic interpretation \cite{Oshita:2016oqn} since it can be shown to be completely determined by the gravitational entropy of the system. Another characteristic of the HM transition is its relatively small decay rate given by \cite{hawking1982supercooled,Weinberg:2006pc}
\begin{equation} \Gamma\lesssim H^4 \exp\left(\frac{1}{V_{\text{i}}}-\frac{1}{V_{\text{top}}}\right)\sim  \Lambda^2\exp\left(-\frac{\delta \Lambda}{\Lambda^2}\right),\label{jeje}\end{equation}
where in the last step we have used that the height of the potential barrier $\delta \Lambda\equiv V_{top}-V_i$ is small compared to the initial vacuum energy $\Lambda=V_i\simeq V_{top}$. Otherwise, the transition will be dominated by thin-walls. This decay rate corresponds to a transition time which is greater than the Hubble time.  This suggests that a HM transition in which the physics does not change drastically  can only be marginally consistent with TCC, since it might still allow for an originally subplanckian mode of the first vacuum to become Hubble-sized after the transition. One might attempt to resolve this tension by requiring the physics in the two vacua to be sufficiently different so the fluctuations in one can no longer be expressed in terms of long-wavelength fluctuations of light degrees of freedom in the other. The distance conjecture suggests that for this to be true, the field must traverse a trans-Planckian range. In the following we show that even trans-Planckian field ranges do not mitigate the tension between HM transition and TCC.

\begin{figure}[t]
\begin{center}

\tikzset{every picture/.style={line width=0.75pt}} 

\begin{tikzpicture}[x=0.75pt,y=0.75pt,yscale=-1.3,xscale=1.3]

\draw  (257,204) -- (464,204)(297,12) -- (297,224) (457,199) -- (464,204) -- (457,209) (292,19) -- (297,12) -- (302,19)  ;
\draw [line width=1.5]    (343,68) .. controls (405,-22) and (396,192) .. (430,173) ;
\draw [color={rgb, 255:red, 74; green, 144; blue, 226 }  ,draw opacity=1 ] [dash pattern={on 4.5pt off 4.5pt}]  (343,68) -- (343,205) ;
\draw [color={rgb, 255:red, 74; green, 144; blue, 226 }  ,draw opacity=1 ] [dash pattern={on 4.5pt off 4.5pt}]  (430,173) -- (430,204) ;
\draw [color={rgb, 255:red, 74; green, 144; blue, 226 }  ,draw opacity=1 ] [dash pattern={on 4.5pt off 4.5pt}]  (370,45) -- (371,204) ;
\draw [color={rgb, 255:red, 208; green, 2; blue, 27 }  ,draw opacity=1 ] [dash pattern={on 4.5pt off 4.5pt}]  (298,68) -- (343,68) ;
\draw [color={rgb, 255:red, 208; green, 2; blue, 27 }  ,draw opacity=1 ] [dash pattern={on 4.5pt off 4.5pt}]  (297,45) -- (370,45) ;
\draw [color={rgb, 255:red, 208; green, 2; blue, 27 }  ,draw opacity=1 ] [dash pattern={on 4.5pt off 4.5pt}]  (297,172) -- (430,173) ;
\draw [color={rgb, 255:red, 155; green, 155; blue, 155 }  ,draw opacity=1 ][line width=0.75]    (343,59) .. controls (345.35,59.24) and (346.4,60.53) .. (346.15,62.88) .. controls (345.91,65.23) and (346.96,66.52) .. (349.31,66.76) .. controls (351.66,67) and (352.71,68.29) .. (352.46,70.64) .. controls (352.22,72.99) and (353.27,74.28) .. (355.62,74.52) .. controls (357.97,74.76) and (359.02,76.05) .. (358.77,78.4) .. controls (358.53,80.75) and (359.58,82.04) .. (361.93,82.28) .. controls (364.28,82.52) and (365.33,83.81) .. (365.08,86.16) .. controls (364.83,88.51) and (365.88,89.8) .. (368.23,90.04) .. controls (370.58,90.28) and (371.63,91.57) .. (371.39,93.92) .. controls (371.15,96.26) and (372.2,97.55) .. (374.54,97.79) .. controls (376.89,98.03) and (377.94,99.32) .. (377.7,101.67) .. controls (377.45,104.02) and (378.5,105.31) .. (380.85,105.55) .. controls (383.2,105.79) and (384.25,107.08) .. (384.01,109.43) .. controls (383.76,111.78) and (384.81,113.07) .. (387.16,113.31) .. controls (389.51,113.55) and (390.56,114.84) .. (390.31,117.19) .. controls (390.07,119.54) and (391.12,120.83) .. (393.47,121.07) .. controls (395.82,121.31) and (396.87,122.6) .. (396.62,124.95) .. controls (396.38,127.3) and (397.43,128.59) .. (399.78,128.83) .. controls (402.13,129.07) and (403.18,130.36) .. (402.93,132.71) .. controls (402.69,135.06) and (403.74,136.35) .. (406.09,136.59) .. controls (408.44,136.83) and (409.49,138.12) .. (409.24,140.47) .. controls (409,142.82) and (410.05,144.11) .. (412.4,144.35) .. controls (414.75,144.59) and (415.8,145.88) .. (415.55,148.23) .. controls (415.3,150.58) and (416.35,151.87) .. (418.7,152.11) .. controls (421.05,152.35) and (422.1,153.64) .. (421.86,155.99) -- (423.06,157.47) -- (428.11,163.67) ;
\draw [shift={(430,166)}, rotate = 230.89] [fill={rgb, 255:red, 155; green, 155; blue, 155 }  ,fill opacity=1 ][line width=0.08]  [draw opacity=0] (5.36,-2.57) -- (0,0) -- (5.36,2.57) -- cycle    ;

\draw (364,206.4) node [anchor=north west][inner sep=0.75pt]    {$\varphi _{c}$};
\draw (335,205.4) node [anchor=north west][inner sep=0.75pt]    {$\varphi _{i}$};
\draw (424,206.4) node [anchor=north west][inner sep=0.75pt]    {$\varphi _{f}$};
\draw (267,58.4) node [anchor=north west][inner sep=0.75pt]    {$\Lambda $};
\draw (240,162.4) node [anchor=north west][inner sep=0.75pt]    {$\Lambda -\Delta \Lambda $};
\draw (242,36.4) node [anchor=north west][inner sep=0.75pt]    {$\Lambda +\delta \Lambda $};

\end{tikzpicture}

\caption{}
\label{HMT}
\end{center}
\end{figure}

We consider the potentials of the form shown in figure \ref{HMT}. Suppose $\Delta \Lambda$ is the energy difference between the initial and final vacua and $\delta\Lambda$ is the height of the potential as shown in figure \ref{HMT}.

Under some circumstances, the TCC implies the refined dS conjecture up to some logarithmic corrections which can be neglected for order of magnitude analysis. We will come back to the required condition and check them later. For now, we assume the refined dS conjecture is true,
\begin{align}\label{ref}
    \frac{|V''|}{V}>\mathcal{O}(1).
\end{align}
If we estimate the potential interpolating between the two vacua with an inverted parabola, we find
\begin{align}\label{ipp}
    \varphi_f-\varphi_i\simeq\frac{\sqrt{2}(\sqrt{\delta\Lambda+\Delta\Lambda}+\sqrt{\delta\Lambda})}{\sqrt{|V''|}}.
\end{align}
As we discussed in the previous section, $\Delta\Lambda\lesssim \delta\lambda$ corresponds to the thin-wall approximation. Therefore, for Hawking-Moss transition we need $\Delta\Lambda\gtrsim \delta\lambda$. Using this inequality we can simplify \eqref{ipp} to find  
\begin{align}\label{ipp2}
    \varphi_f-\varphi_i\sim\sqrt\frac{\Delta\Lambda}{{|V''|}}.
\end{align}
Plugging this into \eqref{ref} leads to
\begin{align}
    \Delta\phi=\varphi_c-\varphi_i<\varphi_f-\varphi_i\lesssim\sqrt\frac{\Delta\Lambda}{\Lambda},
\end{align}
and from $\Delta\Lambda<\Lambda$, we find
\begin{align}
    \Delta\phi\leq \mathcal{O}(1).
\end{align}
Thus, for the potential to be consistent with the TCC the field range must be sub-Planckian, but as we discussed in the beginning of the section, this poses a tension with the other swampland conjectures, in particular the Distance Conjecture.  

Now we go back and check the assumptions we made to get \eqref{ref}. In \cite{bedroya2019trans} it was shown that in $d$ spacetime dimensions, the TCC would imply the refined dS conjecture if
\begin{align}\label{cond}
\Delta\phi\geq \frac{B_1(d)B_2(d)^\frac{3}{4}V_{max}^\frac{d-1}{4}V_{min}^\frac{3}{4}\ln(\frac{B_3(d)}{\sqrt{V_{min}}})^\frac{1}{2}}{V_{min}B_2(d)-|V''|_{max}\ln(\frac{B_3(d)}{\sqrt{V_{min}}})^2},
\end{align}
where $V_{max}$ and $V_{min}$ are respectively the maximum and the minimum of $V$ over $\phi\in[\phi_i,\phi_c]$, and $B_1(d)$, $B_2(d)$, and $B_3(d)$ are $\mathcal{O}(1)$ numbers given by
\begin{align}
B_1(d)=&\frac{\Gamma(\frac{d+1}{2})^\frac{1}{2}2^{1+\frac{d}{4}}}{\pi^\frac{d-1}{4}((d-1)(d-2))^\frac{d-1}{4}},\nonumber\\
B_2(d)=&\frac{4}{(d-1)(d-2)},\nonumber\\
B_3(d)=&\sqrt\frac{(d-1)(d-2)}{2}.
\end{align}
We show that either the above condition holds and henceforth \eqref{ref} is true, or the field range is sub-Planckian. Since we proved \eqref{ref} leads to a sub-Planckian field range, this would prove that in either case the field range must be sub-Planckian which is our final desired result. We prove this claim by contradiction. Suppose $\frac{|V''|}{V}<<\mathcal{O}(1)$ and $\Delta\phi$ is trans-Planckian, in particular $\Delta\phi>>\Lambda^\frac{1}{2}$. We show that this leads to a contradiction.

 Since $|V''|<<V$ the denominator of RHS in \eqref{cond} is dominated by $V$. Moreover, $\delta \Lambda\lesssim\Delta\Lambda<\Lambda$, thus $V_{\max}\sim V_{\min}\sim\Lambda$. Plugging in $V_{max}\simeq V_{min}\simeq\Lambda$ and neglecting the $\mathcal{O}(1)$ terms, including the logarithmic terms, makes \eqref{cond} take the following form.
\begin{align}
    \Delta \phi\lesssim\Lambda^\frac{1}{2},
\end{align}
which is in contradiction with $\Delta\phi>\mathcal{O}(1)$. This proves our claim by contradiction. To summarize, we showed that for the Hawking-Moss transition to be consistent with the TCC the field range traversed during the transition must be sub-Planckian. But then, there is no reason to expect the physics to drastically change after the HM transition, and we run into the problems explained at the beginning of the section: the lifetime associated with HM is of order or greater than Hubble, which might exceed the TCC time getting into tension with the conjecture.

Note that we assumed that second-order expansion around the peak of the potential reasonably approximates the ridge of the potential. One could argue that this makes our derivation somewhat model-dependent.

\bibliographystyle{jhep}
\bibliography{dSrefs}
\end{document}